\documentclass{JHEP3}

\usepackage{amsmath}
\usepackage{latexsym}
\usepackage{graphicx}
\usepackage{bm}
\usepackage{mathrsfs}

\allowdisplaybreaks[1]

\numberwithin{equation}{section}

\def\ZZ{{\mathbb Z}}

\def\be{\begin{equation}}
\def\ee{\end{equation}}
\def\ba{\begin{eqnarray}}
\def\ea{\end{eqnarray}}
\def\hLambda{\hat \Lambda}

\newcommand{\bea}{\begin{eqnarray}}
\newcommand{\eea}{\end{eqnarray}}

\def\al{{\alpha'}}
\def\ie{{\it i.e.}}
\def\threeh{{\scriptstyle {3 \over 2}}}
\def\fiveh{{\scriptstyle {5 \over 2}}}

\def\half{{\scriptstyle {1 \over 2}}}

\def\nn{\nonumber}

\def\q{q}
\def\w{w}
\def\v{v}

\def\a{\alpha }

\def\s{\sigma }
\def\non{\nonumber }
\def\hR{{ R}}

\def\hK{\hat K}

\def\calV{{\cal V}}

\def\calE{{\cal E}}

\title{Non-renormalisation Conditions in Type II String Theory and Maximal Supergravity}
\author{Michael B. Green\\
 Department of Applied Mathematics and
Theoretical Physics\\
Wilberforce Road, Cambridge CB3 0WA, UK\\
\email{\tt M.B.Green@damtp.cam.ac.uk}}
\author{Jorge G. Russo\\
Instituci\' o Catalana de Recerca i Estudis Avan\c{c}ats (ICREA),\\
University de Barcelona,  Facultat de Fisica\\
 Av. Diagonal, 647,  Barcelona 08028 SPAIN\\
\email{\tt jrusso@ub.edu}}
\author{Pierre Vanhove\\
Service de Physique Th{\'e}orique,\\
CEA/DSM/PhT, CEA/Saclay, Orme des Merisiers, CEA/Saclay\\
91191 Gif-sur-Yvette Cedex, France\\
\email{ pierre.vanhove@cea.fr}}

\abstract{ This paper considers general features of the derivative expansion of Feynman diagram
 contributions to the four-graviton scattering amplitude
 in eleven-dimensional supergravity compactified on a two-torus.  These
are  translated into statements about interactions of the form $D^{2k} R^4$ in type II superstring
 theories, assuming the standard M-theory/string theory duality relationships, which provide
 powerful constraints on the effective interactions.
In the ten-dimensional IIA limit we find that there can be
no perturbative contributions beyond  $k$ string loops
 (for $k>0$).  Furthermore, the genus $h=k$ contributions are determined
 exactly by the one-loop eleven-dimensional supergravity amplitude for all values
 of $k$.    A plausible interpretation of these observations is that the sum of $h$-loop
 Feynman diagrams of maximally extended supergravity is less divergent than might be expected
 and could be ultraviolet finite in dimensions $d <  4 + 6/h$ --
the same bound as for $N=4$ Yang--Mills.}

\preprint{hep-th/0610299\\
DAMTP-2006-102\\
SPHT06/127\\
UB-ECM-PF-06-30}
\keywords{Effective action, Superstring}

\begin{document}

\section{Introduction and overview}

Although the complete non-perturbative description of string theory remains
unfathomable, a variety of non-perturbative features of the
derivative expansion of the string theory
effective action have been deduced over the years.
This expresses the low energy dynamics in terms of the massless
fields of the theory after the massive modes have been integrated out.
Considerations have typically been limited to a few higher-derivative
terms (i.e., a few powers of $\alpha'$)
describing a sub-sector of the theory.  A most interesting open
question is to what extent supersymmetry, together with various dualities
might constrain the non-perturbative structure of the terms in the effective
action.  In this paper we will uncover some systematic properties of
the IIA and IIB string effective actions that follow from general
features of the Feynman diagrams of eleven-dimensional supergravity
combined with constraints that enforce the duality relationships between M-theory
and string theory at the quantum level.

At the classical level,  $S$-duality identifies eleven-dimensional
supergravity compactified on a two-torus with classical type IIA or type IIB supergravity
compactified to nine dimensions on a circle \cite{Witten:1995ex,Schwarz:1995jq,Aspinwall:1995fw}.
This identification may be  extended to the
quantum theory by considering the compactification of loop diagrams
of eleven-dimensional supergravity \cite{ggv:oneloop,gv:D6R4,Russo:1997mk,gkv:twoloop}.
 We will here argue that if one assumes that
the duality conditions continue to hold at higher orders, strong constraints are imposed on the
possible higher-genus corrections to higher-derivative terms in the
ten-dimensional type II
effective actions.  We will find, in particular, that the IIA supergravity genus expansion must
satisfy a strong non-renormalisation condition that makes it much less ultraviolet divergent than
would otherwise be suspected.  In fact, our work indicates that, after reduction to $d$ dimensions,
the sum of all diagrams with $h$ loops is  finite in dimensions $d< 4+6/h$.

There is obviously much more to M-theory than the Feynman diagrams of eleven-dimensional supergravity,
which are based on the dynamics of the superparticle and do not explicitly include deeper aspects of M-theory
associated, for example, with $M$-branes.   The non-renormalisability of supergravity perturbation theory
is a symptom that
it does not adequately take short-distance physics into account. Our procedure
in the following will be to introduce an ultraviolet cut-off and subtract divergent terms with counterterms
that have unknown coefficients that parameterize our lack of knowledge of short-distance effects that might
be  determined from first principles in a more fundamental formulation of M-theory.
We will then impose conditions on the expressions that are required by
 duality relationships for consistency with
known properties of string theory.

The structure of the paper is as follows.
 In section~\ref{sec:elevendim} we will consider the general structure of the four-graviton
 amplitude obtained by compactifying the Feynman diagrams of eleven-dimensional supergravity
 on a two-torus of volume $\calV$ and complex structure $\Omega$.
   Our considerations will be restricted to properties of
$L$-loop Feynman diagrams on a two-torus that do not depend on detailed
analysis of the individual diagrams.   We are imagining introducing a momentum cut-off $\Lambda$ to
the loop momenta in such diagrams, which contribute to interactions
 starting from terms of the form
$D^{2\beta_L} R^4$,
where
$\beta_L$ is an integer. As we will briefly review in section~\ref{subsec:comments},
summing all the Feynman diagrams that
contribute to the four-graviton amplitude is known to result in very simple expressions for $L=1$ with  $\beta_1=0$,
and $L=2$ with $\beta_2=2$.
The full dilaton dependence of several low-order terms in the type IIA and IIB string theory derivative expansions
have been obtained in explicit
detail from the toroidally compactified  $L=1$  and $L=2$ amplitudes, making use of the standard duality relation between
M-theory and type II string theory.  These examples, which provide lessons for the rest of the paper, are reviewed in
section~\ref{subsec:lessons}.

General features of four-graviton $L$-loop Feynman diagrams compactified on a torus will be described
in section~\ref{subsec:lloop}.
The superficial degree of divergence of the $L$-loop diagram is
$\Lambda^{9L-6 -2\beta_L}$ (where $9L-6 - 2\beta_L>0$).
However, this power is reduced by $\Lambda^{-w}$ in subdivergent terms that have compensating factors of
$\calV^{-w/2}$ with integer $w>0$.   Furthermore, the four-graviton amplitude has an expansion in powers of the
Mandelstam invariants of the form\footnote{We are using capital
letters $S$, $T$, $U$ for the eleven-dimensional
Mandelstam invariants and lower-case letters for the string theory invariants.} $(\calV S)^\v$,
corresponding to an infinite series of derivatives, $D^{2k}R^4$, where $k=v+\beta_L$.
Invariance under large diffeomorphisms of the torus implies that dependence on the
complex structure
$\Omega$ is necessarily encoded in a $SL(2,Z)$-invariant function.

In subsection~\ref{sec:typetwo} the compactified eleven-dimensional action is translated into
string variables.  The well-known duality relations between M-theory and string theory are used
to make appropriate identifications of the
parameters $\calV$ and $\Omega$ with the moduli of IIA or IIB string theory compactified on a circle
(of radius $r_A$ and $r_B=1/r_A$, respectively).
This leads to a generic description  of the coefficients of the $D^{2k}R^4$ interactions in which
there are  powers of the radius, $r_A^{1+p}$ or $r_B^{1+p}$ (where $p$ depends on $w$ and $k$),
and dependence on the coupling, $e^{\phi}$
(where $\phi$ is the IIA or IIB dilaton) and Ramond--Ramond pseudoscalar (for IIB)
or vector (for IIA).  Imposing these duality rules leads to powerful constraints
on quantities that are undetermined by our
eleven-dimensional supergravity starting point.   For
example, we need to stipulate that the string expansions are in even powers of the string
coupling constant $e^{2\phi}$ and that the most singular term is the tree-level term of
order $e^{-2\phi}$.  When reinterpreted in terms of the eleven-dimensional theory
these conditions severely restrict the possible terms in the effective action.

The treatment of the type IIA and type IIB cases is rather different.    In section~\ref{sec:twob}
we consider the IIB parametrization, where the complex structure $\Omega$ is identified with the
complex coupling constant and the coupling constant dependence of $D^{2k}R^4$ is encoded in
modular functions, $f_{(q,k)}(\Omega,\bar \Omega)$ (where $q=w/2-v$).
Terms that are finite in the ten-dimensional
limit have $p=0$ and the value of $q$ is determined by $k$.  Little can be said about the precise
form of the  modular function $f_{(q,k)}$ without detailed calculations,
such as those that have been determined previously for
$k=0,1,2,3$, which are solutions of Poisson equations in moduli space.  There are also infinite
classes of terms with $p>0$ that diverge in the ten-dimensional limit ($r_B\to \infty$).
 We argue that these terms arise from the low energy  expansion of multi-particle
 threshold singularities for massive Kaluza--Klein modes.   These modes condense to zero mass in the
 ten-dimensional limit and the series of positive powers of $r_B$ must resum to generate the
logarithmic threshold terms that are known to satisfy the ten-dimensional unitarity constraints.
This effect was discussed for the IIA  $L=1$ case  in \cite{gkv:twoloop}.
We will demonstrate how this works for the first few threshold terms in the IIB case.

In the type IIA case considered in section~\ref{sec:twoa} the ten-dimensional limit appears
in a very different
manner.  Unlike the IIB case, the value of $q$ for terms that are finite in ten dimensions
 depends on both $k$ and the string loop genus $h$.  Different orders in the string loop
 expansion arise from
 different values of $L$.  This leads to some surprisingly strong constraints.
 In particular, in the ten-dimensional limit we find that there are no contributions to $D^{2k} R^4$
 beyond $h=k$ loops ($h\ge 1$).
Furthermore, the $h=k$ contributions
 are given exactly by the compactification of one-loop ($L=1$) supergravity for all values of $k>0$.
Terms with $h<k$ can get contributions from arbitrary large values of $L$.

As mentioned earlier our procedure is based on assuming the familiar M-theory/string theory duality
relations, which imposes constraints that are not apparent in ordinary supergravity.
In section~\ref{sec:Mtheory} we will see how the consistency of this procedure restricts the kind of terms
that arise when the theory is lifted back to eleven dimensions.
Since these arguments are based largely on power counting and on the relation
between type II string theory and M-theory parameters
they suggest that similar non-renormalisation conditions might also apply the many other interactions
of the same dimension as $D^{2k} R^4$.

The non-renormalisation conditions of section~\ref{sec:twoafin} provide very powerful restrictions on the
ultraviolet behaviour of low energy IIA supergravity.
 The consequences of these restrictions will be discussed in detail in section~\ref{sec:discuss}.  There
are tantalizing hints that the consistency of the picture presented requires the supergravity theory
to be considerably more finite than is evident from the superficial degree of divergence of individual
Feynman diagrams.
Our results imply that at each successive loop there is an extra factor of $D^2$. In other words,
the sum of all diagrams with $h$ loops has a factor
of $D^{2h}R^4$ so that the degree of divergence is reduced
to that of $\varphi^4$ scalar field theory or electromagnetism coupled to $\varphi^3$
 scalar fields.   If we assume that this feature survives after compactification
this is just what is needed in order for the Feynman diagrams of
$N=8$ supergravity in four dimensions to be ultraviolet finite.  However, care should be
taken in interpreting this low energy limit of string theory since it is one in which infinite towers of
`non-perturbative' states become massless.  This may cast doubt on
the usefulness of the perturbative supergravity approximation, which only includes
the massless perturbative states (similar observations have also
been made by H.  Ooguri and J.H. Schwarz (private communication)).

\section{Eleven-dimensional supergravity compactified on a two-torus}
\label{sec:elevendim}

The description of the effective low-energy theory obtained from eleven dimensions
depends sensitively on the spectrum of massless fields, which
are responsible for branch cuts in amplitudes, and hence for nonlocal terms in the action.
Since the spectrum of massless fields
 is generally a function of the moduli, the form of the effective
 action is different in different patches of moduli space.
 Here we will be concerned with  asymptotically flat nine, ten
and eleven-dimensional space-time.  Thus, starting in nine dimensions,
infinite numbers of Kaluza--Klein modes become massless at the boundaries of  moduli space where
one or two compact dimensions decompactify.  This condensate generates a change in the threshold
behaviour of amplitudes or, equivalently,  generates non-local terms in the higher-dimensional
actions.  Although we will often  talk in terms of the effective action, this can be viewed as a useful shorthand
for encoding the properties of the Feynman diagrams that we are considering.
Properties of the expansion of the amplitude in terms of analytic and non-analytic functions of
Mandelstam invariants translate into properties of local and nonlocal
terms in the  effective action.

The derivative expansion of the local terms in the effective action
is an expansion in powers of $\al=l_s^2$ (where $l_s$ is the string length scale) of the form
\be
\al^4 S= S^{(0)} + \al^3 S^{(3)} + \al^5 S^{(5)} + \al^6 S^{(6)} + \dots\, ,
\label{derivexpan}
\ee
where $S^{(n)}$ denotes the contribution to the action with dimension $2n-8$.
The known properties of the interactions contained in $S^{(n)}$ are
based on a combination of perturbative string theory and the constraints imposed by
dualities and supersymmetry.
Our explicit considerations concern the expansion of the four-graviton amplitude and so they
will be restricted to the linearized approximation to local interactions of the form
\be
R^4\,, \, D^4R^4\,, \, D ^6 R^4\,,\dots\, , D^{2k}R^4\, , \dots,
\label{genform}
\ee
although many of our general  considerations
could also apply to other interactions of the same dimension, such as $R^{k+4}$.
The expression $D^{2k}\, \hR^4$ is shorthand for the particular contractions of tensor indices
 on the linearized curvature
tensors and derivatives, which will not be relevant in this
paper\footnote{For example, at tree-level these contractions  can be extracted from the
  expressions in
appendix~\ref{sec:treelevel}.  For example, the $D^4\, \hR^4$
 term has the form $t_8 t_8\,
R^2\, D^\mu D_\mu\, R^2$, where $t_8$ is the familiar eighth-rank
tensor with indices that contract into pairs of indices from each of the $R$'s.}.
Any possible $D^2 \hR^4$ interaction vanishes on shell and does
not contribute to the four-graviton amplitude.

\subsection{Comments on the structure of eleven-dimensional Feynman diagrams}
\label{subsec:comments}

We will begin by summarizing what is known in general about the structure of the Feynman diagrams
of eleven-dimensional supergravity.

\begin{figure}[h]
\centering
\includegraphics[width=15cm]{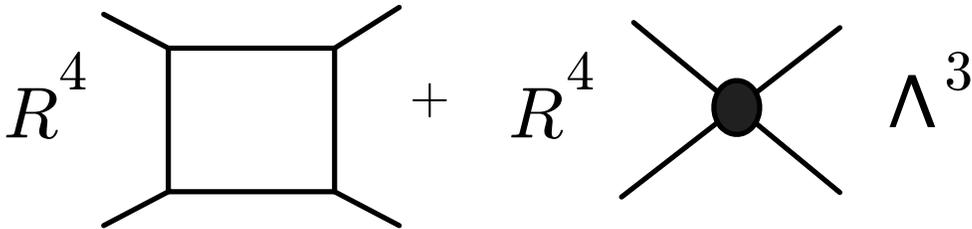}
\caption{The scalar field theory box diagram and the counter-term that subtracts the $\Lambda^3$ divergence.}
\label{fig:BoxDiagram}\end{figure}

It has been known for a long
time that the sum of the
one-loop  ($L=1$) contributions to four-graviton scattering has the
structure of an eleven-dimensional $\varphi^3$ scalar field theory box diagram multiplying $\hR^4$
 (see figure~\ref{fig:BoxDiagram}).   In space-time
dimensions $d\ge 8$ the box is divergent and so it needs to be regularized.  It is cubically
divergent ($\Lambda^3$) in eleven dimensions.   We will subtract this divergence by introducing a counterterm
as illustrated in the figure.
The fact that all the non-trivial dependence
on the external momenta is contained in a scalar field theory box diagram makes this amplitude very
easy to evaluate.

\begin{figure}[h]
\centering
\includegraphics[width=15cm]{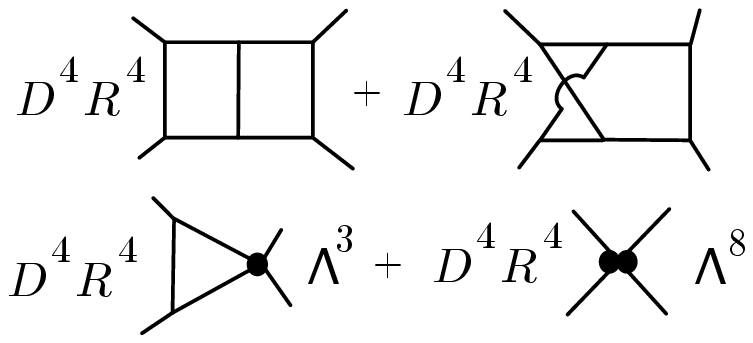}
\caption{The two-loop four-graviton amplitude in eleven dimensions
is  given by the sum of scalar field theory double-box diagrams and counterterms that subtract
the primitive divergence and subdivergences.}
\label{fig:TwoBoxDiagram}\end{figure}

The two-loop ($L=2$) contributions to four-graviton scattering also sum together in a remarkably simple
manner.  The results of \cite{Bern:1998ug} show that
the sum  has an explicit overall factor of the form $S^2\hR^4$ multiplying two particular two-loop Feynman ladder diagrams
of $\varphi^3$ scalar field theory, together with their $T$ and $U$ channel
symmetrization (see figure~\ref{fig:TwoBoxDiagram}).  This means that
$D^4 \hR^4$ is the leading low energy contribution of the two-loop supergravity amplitude and the
two-loop scalar amplitude is divergent in dimensions $d\ge 7$.
In addition to a primitive divergence that behaves
as $\Lambda^8$ when $d=11$ (for a momentum cut-off $\Lambda$), it also has one-loop sub-divergences
that are proportional to $\Lambda^3$ if $d=11$.  The contribution with a single black blob in
figure~\ref{fig:TwoBoxDiagram} is a diagram with the insertion of the one-loop counterterm, which subtracts a sub-divergence.
The double-blob indicates the counterterm for the primitive divergence.  Whereas the $\Lambda^3$ subdivergent contributions
fit in perfectly with string theory expectations (see section~\ref{subsec:lessons}),
the two-loop $\Lambda^8$ contribution is inconsistent
with string perturbation theory, so the coefficient of its counterterm must be chosen so that the contribution of this
term vanishes \cite{gkv:twoloop}.

There has been no complete analysis of properties of the sums of diagrams for loop diagrams with $L>2$.  In
particular, it is not known whether extra factors of $S$, $T$ and $U$ factor out of the
higher-loop amplitudes.    What is known is that there are at least two powers of the
invariants in the prefactor for the sum of loops for any $L\ge 2$, so in the following we will  assume that the
sum of $L$-loop Feynman diagrams has an overall factor of  $D^{2\beta_L} \hR^4$,
where $\beta_L \ge 2$.   This allows for the possibility
that extra derivatives might be extracted at $L$-loops although our arguments do not assume this.

In the absence of a simple description of the sum of Feynman diagrams when $L>2$ we will not be able to evaluate the
higher loop effects explicitly and our discussion will be based on general
issues, making extensive use of dimensional analysis.
Wherever a power of the cut-off
appears it signifies divergences that are canceled by counterterms and the result is finite or zero.
In the following, when we use the symbol $\Lambda$ it will usually be with the understanding that the
cut-off has been canceled by a counterterm so that $\Lambda^m$ denotes a finite
but undetermined constant with dimensions $l_P^{-m}$, where $l_P$ is the eleven-dimensional Planck
distance.

\subsection{Lessons from  known higher derivative terms in the IIA and IIB actions}
\label{subsec:lessons}

In order to illustrate the general structure we will first review in more detail the explicit
description of certain string theory effective interactions that follow from
one-loop ($L=1$) and two-loop ($L=2$) eleven-dimensional supergravity outlined in the last subsection.
 This has lead to an
understanding of the dilaton dependence of interactions up to $\al^2\, D^6 \hR^4$, which we will review
in this subsection.

The first term beyond the classical supergravity action, $S^{(0)}$
(the supersymmetric
completion of the $O(\al^{-4})$ Einstein--Hilbert action) is $S^{(3)}$,
which includes the well-known  $\hR^4$ interaction.
In the IIB theory this is given by \cite{gg:dinstantons}
\be
\label{rfour}
S^{(3)} \sim
\int d^{10}x\sqrt{-g}\, \Omega_2^{1\over2}\,
Z_{\threeh}(\Omega,\bar\Omega)\, R^4 ,
\ee
where $\Omega= \Omega_1 + i\Omega_2$
is the complex scalar field of the type IIB theory ($\Omega_{2}=\exp(-\phi^B)$ is the type IIB coupling constant) and
$Z_{3/2}(\Omega,\bar\Omega)$ is a $SL(2,Z)$-invariant function.  Various arguments have
established that $Z_{3/2}$ is a non-holomorphic Eisenstein series,  which is the
$s=3/2$ case of the more general series,
\be
\label{gene}
Z_s = \sum_{(m,n)\ne (0,0)} \frac{\Omega_2^s}{|m+ n\Omega|^{2s}}\, .
\ee
In addition to (\ref{rfour}) there are many other interactions of the same
dimension that are related to $R^4$ by the classical supersymmetries.  These in general
involve combinations of fields that transform as $(-w,w)$ forms  under the $SL(2,Z)$ duality group \cite{Green:1998by}
(where the notation indicates  holomorphic and anti-holomorphic weights).
The coefficient functions that generalize $Z_{3/2}$ are $(w,-w)$-forms, $Z_{3/2}^{(w,-w)}$.
The full set of such interactions
has not been enumerated,  although they are known in linearized approximation.
The Eisenstein series, $Z_s$ is the unique solution of a Laplace eigenvalue equation
on the fundamental domain of $SL(2,Z)$ with a polynomial growth in $\Omega_{2}$ at infinity,
\be
\Delta_{\Omega} Z_s = s(s-1) Z_s\, .
\label{lapeig}
\ee
Expanding (\ref{gene}) for large $\Omega_2$
(\ie, for small string coupling) gives
\ba
Z_s (\Omega,\bar \Omega) & = &2
\zeta(2s) \Omega_2^{s}  + 2\sqrt \pi  \Omega_2^{1 - s }
{\Gamma(s-\half)\zeta(2s -1)\over \Gamma(s)}\nn\\
& & + {2\pi^{s}\over \Gamma(s)}  \sum_{k\ne
0} \mu(k,s)
 e^{-2\pi(|k|\Omega_2 - i k \Omega_1)}
 |k|^{s-1}
 \left(1+{s(s-1) \over 4\pi |k| \Omega_2} +\dots\right)\,,
\label{eexpan}
\ea
where the last term comes from the asymptotic expansion of a modified
Bessel function and $\mu(k,m) = \sum_{d|k}1/d^{2m-1}$.  This expression
has two power-behaved terms, which are identified with perturbative string theory
tree-level and $l$-loop terms,
as well as an infinite number of non-perturbative exponentially suppressed
$D$-instanton contributions.  This demonstrates, for example,
the perturbative non-renormalisation of $R^4$ in the IIB theory beyond one string loop.
The coefficients of the tree-level and one-loop terms coincide with the
known values found directly from perturbative string theory, which are
summarized in appendix~\ref{sec:treelevel} and ~\ref{sec:oneloop}\footnote{
The expansion of the one-loop amplitude
up to the coefficient of $D^6\hR^4$ was considered in  \cite{gv:stringloop}.
Recently, we have extended
this to include the coefficients of the
$D^8\hR^4$, $D^{10}\hR^4$ and $D^{12} \hR^4$ interactions \cite{grv:stringloop}.}.
The IIA theory has no non-trivial duality symmetry.  Its action has
the same perturbative terms as in the IIB case but there are no $D$-instantons, and hence
no non-perturbative contributions.

These properties of the $R^4$ term can be deduced by evaluating the one-loop
contribution to four-graviton scattering in
eleven-dimensional supergravity compactified
to nine dimensions on a two-torus of volume $\calV$ and complex
structure $\Omega$.
The amplitude can be expressed as a sum of windings of the loop around the two
cycles of the torus, in which case the ultraviolet divergence is entirely in the
zero winding sector.
The duality properties of M-theory
imply that the IIB theory is obtained in the limit in which the
volume of the two-torus vanishes, in which case the coefficient of the
ultra-violet divergence also vanishes, giving a finite result  proportional to
$Z_{3/2}(\Omega,\bar \Omega)$.  Furthermore, the IIA theory arises from the
limit in which the torus decompactifies to a circle.  In that case,  there is a finite
contribution proportional to $1/R_{11}^3$ from the sum of terms in which the loop has non-zero winding
around the circle, which is identified with the string theory
tree-level contribution.   The ultra-violet divergence is contained entirely in the zero winding
term and its coefficient is proportional to $\Lambda^3$ (where $\Lambda$ is a momentum cut-off), which needs to be
canceled by adding a counterterm (see figure~\ref{fig:BoxDiagram}).
Since the zero winding sector is independent of $R_{11}$ it
corresponds to one loop ($h=1$) in the IIA string theory.  As shown in \cite{ggv:oneloop},
the precise  renormalised coefficient is fixed by imposing the fact that the IIA and IIB
string theories are known to have identical four-graviton one-loop amplitudes.\footnote{
The equality of IIA and IIB four-graviton perturbative contributions persists to at least
four loops.  The difference between perturbative contributions in IIA and IIB string theories lies in the sign of
the odd-odd spin structures, which are associated with the presence of $\epsilon^{\mu_0\dots \mu_9}$
tensors in both the left-moving and right-moving sectors.
The indices on these tensors can only be saturated
when there are three or more picture-changing operators in the left-moving and right-moving
sectors, but this
requires at least three loops.  However,
at three loops it is also possible to show that the amplitudes must be
equal,
and there is a strong argument for their equality at four loops \cite{Berkovits:2006vc}.}
The value of the the IIB one-loop coefficient contained in $Z_{3/2}$ therefore determines the
one-loop counterterm to take the value
${2\pi^2/3}- \hLambda^3$ where $\hLambda=l_{P}\Lambda$.

The next order of the derivative expansion that contributes to four-graviton scattering
in the IIB theory is an interaction of the form \cite{gkv:twoloop}
\be
S^{(5)} \sim  \,\int d^{10}x\sqrt{-g}\,
\Omega_2^{-\frac{1}{2}}\, Z_{\fiveh}(\Omega,\bar\Omega)\,D^4 \hR^4\ .
\label{nextord}
\ee
The function $Z_{5/2}$ contains perturbative tree-level and two-loop
terms, which have coefficients that agree precisely with those calculated in perturbative string
theory, and there are no further perturbative terms.  The fact that
there is no one-loop
$D^4\hR^4$ contribution is also consistent with  string perturbation theory \cite{gv:stringloop}.
In this case the properties of the $D^4 \hR^4$ interaction can
be deduced by considering  the two-loop
contribution to four-graviton scattering in eleven-dimensional supergravity
compactified to nine dimensions on a two-torus
\cite{gkv:twoloop}.  There is now a pair of winding numbers associated
with each loop.  The $D^4\hR^4$ contribution comes from a sector in
which one loop has zero winding number, which gives a one-loop
subdivergence  proportional to $\Lambda^3$ (see figure~\ref{fig:TwoBoxDiagram}).
Inserting the same one-loop
counterterm as was used in the one-loop $R^4$ calculation leads to (\ref{nextord}).
As before, the IIA theory is obtained
by considering compactification on a circle.  The only non-zero contributions
come from the power-behaved terms, in this case corresponding to tree-level and two-loop
string perturbation theory, and are equal to those of the IIB case.

At the next order in the $\alpha'$ expansion the four-graviton
amplitude displays  significant new features.   This is indicated by the
the tree-level coefficient which is  proportional to $\zeta(3)^2$ (as
reviewed in appendix~\ref{sec:treelevel}).  As shown in \cite{gv:D6R4} the term in
the effective IIB action that contributes to four-graviton scattering  has the form
(in string frame)
\be
S^{(6)}\sim \int d^{10}x\sqrt{-g}\, \Omega_2^{-1}\,
\calE_{(\threeh,\threeh)}(\Omega,\bar\Omega)\, D^6 \hR^4\, ,
\label{thirdord}
\ee
which was obtained by expanding the two-loop
eleven-dimensional supergravity amplitude
to the first non-leading order in Mandelstam invariants.  This $D^6\hR^4$ contribution
arises as a finite contribution to the two-loop amplitude since it is given by a sum over
non-zero windings of the two loops around both the cycles of the two-torus.
The modular function  $\calE_{(3/2,3/2)}(\Omega,\bar\Omega)$  satisfies
the Poisson equation
\be
\Delta \calE_{(\threeh,\threeh)} =
12 \calE_{(\threeh,\threeh)} -6 Z_\threeh\, Z_\threeh\, .
\label{poiss}
\ee
The inhomogeneous source term makes this equation quite different from the Laplace
eigenfunction equation (\ref{lapeig}).  Its
structure was argued in \cite{gv:D6R4} to follow, at
least qualitatively, from the constraints of supersymmetry.
In this case the expansion of
$\calE_{(3/2,3/2)}(\Omega,\bar\Omega)$ contains tree-level, one-loop, two-loop
and three-loop perturbative string theory terms, as well as infinite series of
$D$-instanton and double $D$-instanton terms.  The tree-level and one-loop coefficients
precisely reproduce those known from string calculations.  The two-loop coefficient has not
yet been extracted from the expression for the two-loop string amplitude
in \cite{DhokerPhongTwoloop,DhokerPhongGutperleTwoloop, berkovitstwoloop} so it remains a `prediction'.
Rather remarkably, the
type IIB three-loop coefficient extracted from $\calE_{(3/2,3/2)}$ is exactly the same as
the three-loop contribution to the type IIA theory
that was extracted from the one-loop ($L=1$) supergravity amplitude in
\cite{Russo:1997mk,gkv:twoloop} (and will be reviewed later in this paper) --- which is in agreement
with string theory expectations.

\subsection{The $L$-loop four-graviton amplitude compactified on a two-torus}
\label{subsec:lloop}

We turn now to consider the expression for the $L$-loop contribution to four-graviton scattering
in asymptotically flat eleven-dimensional space-time compactified on a two-torus
of volume $\calV$ and complex structure $\Omega$.  In the following the
 ultraviolet divergences will be regulated by simply introducing a momentum cut-off $\Lambda$,
 as we did for the one-loop and two-loop cases, although
we expect that the precise details of how this is implemented will not be relevant to the following
general arguments\footnote{A simple momentum
cut-off obviously breaks the local symmetries, which would then have to be restored by
the counterterms -- a procedure that would be
very difficult to implement explicitly in practise.}.  To begin with, we
will describe the analytic contributions to the amplitude, which translate into
local terms, $S_L$, in the action.  The derivative expansion will be written
as $S_L = \sum_{\v =0}^\infty S_L^{(\beta_L+3+v)}$,
where $\v$ labels the power of the derivatives in the expansion.    Dimensional analysis determines that
 the leading low energy contribution ($\v=0$) has the form
\be
S^{(\beta_L+3)}_L = l_P^{9(L-1)} \,\sum_{\w=0}^{ \w_{L}}
\Lambda^{9L-6-2\beta_L - \w} \int d^9 x \sqrt{-G^{(9)}}\,
\calV^{1-{\w\over 2}}
\,f_{({\w\over 2},\beta_L)}(\Omega,\bar\Omega)\,  D^{2\beta_L}\hR^4\, ,
\label{minkelev}
\ee
where $-G^{(9)}$ is the determinant of the nine-dimensional metric and
$l_P$ is the eleven-dimensional Planck distance.
 The function
$f_{(\w/2,\beta_L)}(\Omega,\bar\Omega)$  is invariant under modular transformations
(large diffeomorphisms) of the two-torus and, as noted earlier, direct Feynman diagram
calculations give $\beta_1=0$ and
$\beta_2=2$.  Furthermore the $R^4$ interaction is protected by supersymmetry
from renormalization beyond one loop  (see, for example,
\cite{Berkovits:2006vc}) so  $\beta_L \ge 2$ for $L>2$.
  The integer $\w$ determines the power of $\Lambda$, with $\w=0$ for the leading divergence and
$\w>0$ for subdivergences,
which are associated with corresponding inverse powers of $\calV$
(possible $(\ln \Lambda)^n$ factors are not explicitly shown since
they are not seen in the following power-counting analysis).  The values of $\w$ that are
summed over are bounded by $w_{L}\leq 9L-6-2\beta_{L}$ and
 depend on the nature of these divergences.
As described earlier, a power of $\Lambda$ in (\ref{minkelev}) and  subsequent equations will
be taken to represent the undetermined renormalised value that results when divergences are
canceled by adding appropriate counterterms.
 This value will be assumed to be a constant in Planck units.

Higher order terms in the derivative
expansion are obtained by expanding the amplitude in
powers of the Mandelstam invariants $S$, $T$ and $U$.  Dimensional analysis gives
an expression for $S_L$ as an infinite series of powers of $\calV D^2$ in which the
general term $(\v \ge 0$) has the form
\ba
\label{twotorus}
S_L^{(\beta_L+\v +3)}  &=& \sum_{\w=0}^{w_{L}}
l_P^{9(L-1)} \, \Lambda^{9L-6-2\beta_L -\w}\nn\\
&& \qquad \int d^9 x \sqrt{-G^{(9)}}\, \calV^{1-{\w\over 2}} \,
f_{({\w\over 2}-\v,\beta_L+\v)}(\Omega, \bar\Omega)(\calV D^2)^\v \, D^{2\beta_L}\hR^4\, ,
\ea
 We may now define
\be
k=\v+\beta_L\, \qquad \q  = {\w\over 2} -\v\, .
\label{ksdef}
\ee
and write (\ref{twotorus}) as
\be
\label{twotorusnew}
S_L^{(k+3)} = \sum_{\q=  \beta_{L}-k}^{q_L}
l_P^{9(L-1)} \, \Lambda^{9L-6-2k-2\q} \int d^9 x \sqrt{-G^{(9)}}\, \calV^{1-\q } \,
f_{(\q ,k)}(\Omega, \bar\Omega)\,  D^{2k}\hR^4\, .
\ee
Note that $k\ge \beta_L \ge 0$, while $\q $ can be of either sign and has the upper limit
$\q_L = \beta_L-k+\w_L/2$.
Equation (\ref{twotorusnew})
involves a sum of terms with different values of $\q $ that depends on details of which
values of $\w$ arise in the expressions for the $L$-loop amplitude.  Terms with positive powers of $\q $
are suppressed in the large-$\calV$ limit while those with negative powers diverge as $\calV
\to \infty$.  This distinctive behaviour will be discussed further below.

The modular function $f_{(\q,k)}$ is undetermined by our general analysis.
However, duality with string theory requires that it has an expansion in powers
of $\Omega_2^{-2}$ in order to  correspond to the string
perturbation expansion in powers of  $g_B^2$, together with an infinite series
of non-perturbative $D$-instanton terms  that has the general form
\be
\label{omexpand}
f_{(\q ,k)}(\Omega,\bar\Omega) = c^{\q ,k}_0 \Omega_2^{a_{\q ,k}} + c^{\q ,k}_1 \Omega_2^{a_{\q ,k}-2} + \dots +
c^{\q ,k}_h \Omega_2^{a_{\q ,k} -2h}
+\dots + \sum_{k\ne 0, \, l\ge |k|} b^{\q ,k}_{k,l} \, e^{2\pi i k\Omega_1} e^{-2\pi l \Omega_2}\,.
\ee
The exact values of the coefficients and powers in this expression cannot be determined purely from
perturbative supergravity.  However, as we will shortly see,
 the value of the constant $a_{\q ,k}$ is determined if we assume duality with
 string theory, so that the leading power
of $\Omega_2$ coincides with tree level (genus $0$)
 in IIB string theory, and $h$ is the genus in
the string theory interpretation.

\subsection{Relation to type II string theories.}
\label{sec:typetwo}

In order to discuss the string theory interpretation of the $L$-loop amplitude
we will first review the well-known dictionary
that expresses the correspondence between eleven-dimensional supergravity and
 type II string theories  \cite{Witten:1995ex,Schwarz:1995jq,Aspinwall:1995fw}.
The eleven-dimensional metric is related to the IIA string metric by
\be
\label{toruscomp}
(ds)^2= G_{MN}^{(11)} dx^M dx^N =
{l_{P}^3\over l_{s}^2}\,[ R_{11}^{-1} (ds^{(9)}_{IIA})^2 +R_{11}^{-1} \, G_{ij}dx^idx^j]\, ,
\ee
where $M,N = 0,1,\cdots, 9, 11$ are eleven-dimensional Lorentz vector indices, $ds^{(9)}_{IIA}$ is the
type IIA element of length in nine dimension ($M, N = 0, 1, \dots, 8$) and $i,j = 9,11$.
The torus metric $G_{ij}$ can be written in the form
\be
\label{tormet}
G = \frac{\calV}{\Omega_2}\begin{pmatrix}
 |\Omega|^2 & -\Omega_1 \\
               -\Omega_1 &1
\end{pmatrix}\, ,
\ee
where the complex structure and volume of the
torus are given by $\Omega = \Omega_1+ i \Omega_2$ and  $\calV = R_{9}R_{11}$,
where  $R_{9}$ and $R_{11}$ are
the radii of its cycles.   In the following
we will only consider the special cases in
which the indices on the polarization tensors and momenta are in the non-compact nine-dimensional
directions, $0,1, \dots, 8$.  From this, and noting the
factor of $R_{11}^{-1}$ in front of the nine-dimensional metric on the right-hand side of
(\ref{toruscomp}), it follows that the
eleven-dimensional Mandelstam invariants for four-graviton scattering\footnote{These invariants
are defined by
$S = -\eta^{MN} (k_1+k_2)_M (k_1+k_2)_N$, $T = -\eta^{MN} (k_1+k_2)_M (k_1+k_2)_N$,
$U = -\eta^{MN} (k_1+k_2)_M (k_1+k_2)_N$,
where $k_r$ ($r=1,2,3,4$) are the moment of the four gravitons and $\eta_{MN}$ is the
eleven-dimensional Minkowski metric.} are related to those of string theory
by
\be
l_{P}^2\,S= {R_{11}\over l_{P}}\, l_{s}^2\, s\, , \qquad l_{P}^2\,T={ R_{11}\over l_{P}}\,l_{s}^2\, t\, ,
 \qquad l_{P}^2\,U= {R_{11}\over l_{P}}\,l_{s}^2\, u\,,
\label{invariants}
\ee
where lower case letters denote the string theory invariants in the
string frame.

The parameters of the corresponding type IIB string theory on a circle
of radius $r_B$ are given by
\begin{equation}\label{dictionary}
r_{B}^{-1}= R_{9}R_{11}^{1\over2}\,l_{P}^{-{3\over2}}, \qquad
 C^{(0)}=\Omega_{1},\qquad e^{-\phi_B}= \Omega_{2}= {R_{9}\over R_{11}}\, ,
\end{equation}
where $C^{(0)}$ is the Ramond--Ramond zero-form and $e^{\phi_B}$ is the IIB coupling,
 and $r_{B}$
is the dimensionless length (in string units) of the
 radius of compactification from ten to nine dimensions.
Note that the torus volume is given by
\be
\label{voltor}
{\cal V}=R_{9}R_{11}=e^{\phi_B/3}\,   r_{B}^{-{4\over3}}\,l_{P}^2\, .
\ee
The type IIA theory is obtained using the identifications
\be
 r_A=R_{9}R_{11}^{1\over 2}\, l_{P}^{-{3\over2}}\, ,\qquad C^{(9)} = \Omega_1\,,
 \qquad e^{\phi_A}= R_{11}^{3\over 2}\, l_{P}^{-{3\over2}}\, ,
\label{iiadict}
\ee
where $C^{(9)}$ is the component of the Ramond--Ramond one-form along the compact dimension of
radius $R_{9}$,  $r_{A}$ is the dimensionless length (in string units) of the
 radius of compactification from ten to nine dimensions.  It follows that the IIB parameters are related to those of IIA by
\be
r_B = r_A^{-1}\, , \qquad r_B e^{-\phi_B} =e^{-\phi_A}\, .
\label{relattwo}
\ee

Equation (\ref{twotorusnew})
can easily be rewritten in terms of IIB string theory coordinates in the string frame as
 \be
 \label{sumssB}
S^{(k+3)B}_L = \sum_{\q= \beta_{L}-k}^{\q_L} S^{(k+3)B}_{m,\q }\, ,
 \ee
 where
\be
\label{twobact}
S^{(k+3)B}_{m,\q}=l_{s}^{2k-1}
 \hLambda^{m} \int d^9 x \sqrt{-g_B^{(9)}}\,
r_B^{-1+ {4\q \over 3} - {2k\over 3}}\, e^{({2k\over 3}-{\q \over 3})\phi_B}
 \,
f_{(\q ,k)}(\Omega, \bar\Omega)\,  D^{2k}\hR^4
\,,
\ee
and the complex structure, $\Omega$, is now interpreted as the complex IIB coupling constant.
In this equation we have introduced the dimensionless cut-off $\hLambda= l_P\, \Lambda$ expressed in M-theory units
and the power of the cut-off is denoted by
\begin{equation}\label{mdef}
m=9L-6-2k-2q \,.
\end{equation}
The factor of $e^{(2k/3-\q /3)\phi_B}$ is absent in the Einstein frame.
The perturbative string theory contributions are obtained by substituting
(\ref{omexpand}) into (\ref{twobact}).   In order for the leading power to correspond
to string tree level behaviour,
$\Omega_2^{2} = e^{-2\phi_B}$, we need to set
\be
a_{\q ,k} = 2 +\frac{2k}{3} - \frac{\q }{3}\, .
\label{treebehav}
\ee
Terms proportional to $r_B$, which give finite contributions in the  ten-dimensional limit,
$r_B\to \infty$, are obtained by setting
\be
\label{rblim}
\frac{4\q }{3} - \frac{2k}{3} =2\, ,
\ee
which fixes the value of $\q $ for a given  $k$.
Thus a non-zero $L$-loop eleven-dimensional supergravity contribution to $D^{2k}\hR^4$ in the IIB
ten-dimensional limit  comes from diagrams which have a $\Lambda^{9(L-1)-3k}$ dependence on the cut-off
(for $L>1$, since $9(L-1) -3k$ is negative when $L=1$).

Terms with $4\q /3 - 2k/3 < 2$ vanish in the
large-$r_B$ limit whereas those with  $4\q /3 - 2k/3 > 2$ diverge in the
large-$r_B$ limit.  The
terms that grow with powers of $r_B$ have to
resum in a manner that generates the non-analytic logarithmic thresholds in ten dimension as
discussed for the case of the leading supergravity threshold in \cite{gkv:twoloop}, as
we will see in section~\ref{sec:twoadiverge}.  For terms that are
proportional to $r_B^{1+p}$ the relation (\ref{rblim}) generalizes to
\be
\frac{4\q }{3} - \frac{2k}{3} =2+2p\, .
\label{rbnew}
\ee

The expression (\ref{twotorusnew})
 can also be rewritten in terms of IIA string-frame  coordinates as
 \be
 \label{sumssA}
S^{(k+3)A}_L = \sum_{\q= \beta_{L}-k}^{\q_L} S^{(k+3)A}_{m,\q }\, ,
 \ee
 where
\ba
S^{(k+3)A}_{m,\q } &=& l_s^{2k-1} \,  \hLambda^{m} \int d^9 x \sqrt{-g_A^{(9)}}\,
r_A^{1-\q +a_{\q ,k}} \, e^{( {2k\over 3} -{\q \over 3})\phi_A}
 \, (c_0\, e^{-a_{\q ,k}\phi_A}
 \nn\\
 &+&  c_1\, r_A^{-2}\, e^{-(a_{\q ,k}-2)\phi_A} + \dots +
 c_h\, r_A^{-2h}\, e^{-(a_{\q ,k}-2h)\phi_A} + \dots)\,  D^{2k}\hR^4\, .
\label{twoaactone}
 \ea
In this expression we have exhibited the perturbative terms in
the expansion of $f_{(\q ,k)}(\Omega,\bar\Omega)$ (\ref{omexpand})
(for clarity, we have set $c_i^{q,k}\equiv c_i\ ,\ i=0,...,h$).
 Once again, if we assume that the leading term is the tree-level string theory interaction
 we have $a_{\q ,k} = 2+2k/3 -\q /3$, so that
\ba
S^{(k+3)A}_{m,\q }
 &=&  l_s^{2k-1} \,  \hLambda^{m} \int d^9 x \sqrt{-g_A^{(9)}}\,
r_A^{3-\frac{4\q }{3}+\frac{2k}{3}} \,\nn\\
&& (c_0\, e^{-2\phi_A}
+ c_1\, r_A^{-2} + \dots + c_h\, r_A^{-2h}
\, e^{2(h-1)\phi_A}+ \dots)\,  D^{2k}\hR^4
\,,
\label{twoaacttwo}
\ea
In this case the condition on $\q $ that a given term should have a finite large--$r_A$ limit
requires
\be
\label{largera}
\frac{4\q }{3} - \frac{2k}{3} = 2 -2h\, .
\ee
The value of $\q $ satisfying this equation
depends on the genus, in contrast to the IIB case (\ref{rblim}).   Furthermore, in the limit
$r_A\to \infty$, the $D$-instanton terms in (\ref{omexpand}) vanish (since the $D$-instanton action
is proportional to $r_A$).
More generally, a term behaving as $r_A^{1+p}$ is selected by choosing
\be
\frac{4\q }{3} - \frac{2k}{3} = 2-2h -p
\label{genrterm}
\ee

\section{Higher derivative interactions in type IIB}
\label{sec:twob}

In the absence of further information the
modular functions $f_{(\q ,k)}$  are undetermined and can contain arbitrary powers of  the string
coupling, $\Omega_2^{-1}$.  Therefore,
the general structure of the IIB action in (\ref{twobact}) does not lead to
non-renormalisation statements (of the kind that we will find later
in the IIA case) but a number of interesting
systematic statements can be made.   We will first discuss terms that grow linearly with
$r_B$ and contribute a finite
quantity to the ten-dimensional string effective action.   We will then consider
terms involving higher powers
of $r_B$, which diverge in ten dimensions but can be resummed to give finite, nonlocal contributions
to the ten-dimensional theory that correspond to threshold cuts in the amplitude.

\subsection{Terms that are finite as $r_B\to \infty$}

 The terms proportional to $r_B$ that are finite in the large-$r_B$
 limit have (from (\ref{rblim}))
$\q =3/2 + k/2$, which means that the dilaton prefactor in (\ref{twobact}) is
$e^{(2k-\q )\phi_B/3} = e^{(k - 1)\phi_B/2}$.   This means that the modular functions,
$f_{(\q ,k)}(\Omega,\bar\Omega)$, that contribute to tree-level string scattering have an
expansion that starts with $\Omega_2^{(3+k)\phi_B/2}$ (since  $a_{\q ,k} =(k+3)/2$).
More information is needed
to determine the particular modular functions that
arise for any value of $k$.  For the cases $k=0, 2, 3$ the modular functions,
determined by explicitly compactifying one and two-loop
eleven-dimensional supergravity on a torus, are known to
be $f_{(3/2,0)} = Z_{3/2}$, $f_{(5/2,2)} = Z_{5/2}$, $f_{(3,3)} = \calE_{(3/2,3/2)}$, where the
generalized Eisenstein series satisfy the Laplace or Poisson equations described in the
introduction.  There is impressive agreement between the coefficients of the perturbative terms
contained in these functions and those that have been calculated at tree-level and one-loop
in string theory. There is also internal consistency between the IIA and IIB calculations.
Whereas the $R^4$ term is protected from getting any contribution from $L>1$
supergravity loops, there is no known reason for the $L=2$ calculations of the
$D^4R^4$ and $D^6R^4$ interactions to be protected.  However, the agreement of these coefficients
suggests that there is some as yet undiscovered non-renormalisation condition.

\subsection{Terms that diverge as $r_B\to \infty$}
Apart from terms that are finite or vanish in the large-$r_B$ limit,
(\ref{twobact})
also allows for terms that diverge in the ten-dimensional limit.  Such
terms that are
essential for building up the non-analytic threshold behaviour of the
amplitude in
 ten dimensions \cite{gkv:twoloop}.  The form of these non-analytic terms is highly constrained by
unitarity but this is obviously very difficult to analyze explicitly in the general case.
However, if we restrict our considerations to two-particle thresholds we will be able to
pinpoint some essential features of certain infinite series' of terms.
The $s$-channel two-particle discontinuity of the amplitude in ten dimensions is given by
 the integral over two-particle phase space of the product of two four-graviton
 amplitudes (we are not here concerned with exact coefficients),
\ba
{\rm Disc}_{\rm s}\, A_4(s,t,u) &\sim& \int d^{10}q A_4(k_1, k_2, q, -q-k_1-k_2)
 \, A_4^{\dagger}(k_3, k_4, -q, q+k_3+k_4)\nn\\
&& \qquad \delta^{(10)}(q^2)\, \delta^{(10)}((q+k_1+k_2)^2) \theta(q_{0})\theta((q+k_{1}+k_{2})_{0})\, .
\label{unigen}
\ea
In nine dimensions the integral includes the sum over intermediate states that include
an infinite sequence of massive Kaluza--Klein two-particle states.

The low energy expansion of (\ref{unigen}) involves expanding each factor of $A_4$
in powers of $\alpha'$, so the derivative expansion of the amplitude feeds back into
the expression for the normal thresholds.  We will distinguish the
analytic part of the amplitude, $A^{an}$, from the nonanalytic part,  $A^{nonan}$, that has
singularities, which is associated with thresholds of various kinds.
So we will write
\be
A_4 = A_4^{an} + A_4^{nonan}\, .
\label{ampdiv}
\ee
The $A_4^{an}$ term encodes local terms in the effective action whereas $A^{nonan}$ contains
normal thresholds  that correspond to non-local terms.
 For the type IIB theory in ten dimensions
the expansion of $A_4^{an}$ has the schematic form (that was reviewed
 in the introduction)\footnote{Constant coefficients have been omitted
 and dependence on $t$ and $u$ has been suppressed in this and the
 following equations.}
\be
A^{an}_4 \sim {\alpha'}^{-4}\, \Omega_2^2\, A_4^{Born} + \left( {\alpha'}^{-1} \,\Omega_2^{\half}\, Z_{\threeh} +
 {\alpha'} \,\Omega_2^{-\half}\, Z_{\fiveh}\, s^2
+ {\alpha'}^2 \,\Omega_2^{-1}\, \calE_{(\threeh,\threeh)}\, s^3  + O({\alpha'}^3)\right)\,
\hR^4\, ,
\label{schemaf}
\ee
where $A_4^{Born}$ is the tree-amplitude that corresponds to the Einstein--Hilbert part of the
action.
 Substituting this into (\ref{unigen})
leads to expressions for the first few threshold contributions
to the amplitude.  For the purposes of this section we need only describe a few of the many threshold terms
that arise up to order ${(\alpha'})^7$,
\ba
A_4^{nonan} &\sim& \left(s\ln(-\alpha' s) + {\alpha'}^3\,\Omega_2^{-\threeh}\, Z_{\threeh}\,
s^4 \ln(-\alpha' s) + {\alpha'}^4\,\Omega_2^{-2}\,   s^5 \ln^2(-\alpha' s)\right.
 \nn\\
&&
\left.+ {\alpha'}^5\,\Omega_2^{-\fiveh}\, Z_{\fiveh}\, s^6 \ln(-\alpha' s)+ {\alpha'}^6\,\Omega_2^{-3}\,
(Z_{\threeh}\times Z_{\threeh} + \calE_{(\threeh,\threeh)}) \,  s^7 \ln(-\alpha' s) +\dots\right)\,
\hR^4\, .
\nn\\
\label{twothresh}
\ea
Note that this expression includes a $s^5\ln^2(-\alpha' s)\hR^4$
that is the first contribution that arises by substituting a loop
contribution into one of the factors of $A_4$ inside the integral on
the right-hand side of (\ref{unigen}).  This term has a double
discontinuity and  arises from a two-loop supergravity diagram.

The separation between analytic and nonanalytic terms in (\ref{ampdiv})
is ambiguous
since unitarity does not determine the scale of the argument of the
$\ln(-s)$'s in (\ref{twothresh}).  We have chosen to normalize the
arguments of each logarithm so that it vanishes at the string scale,
$s=1/\alpha'$.  Changing that scale amounts to a shift in the value of
the coefficient of a corresponding analytic term.  For example, $s^4
\ln (-\alpha' s/A) = s^4 \ln (-\alpha' s) - s^4 \ln A$.  In terms of
the effective action this scale defines the scale for the Wilsonian
cut-off, which is an arbitrary parameter in the action.

It is, of course, the complete amplitude that is supposed to be
invariant under duality transformations,  $SL(2,Z)$ for type IIB in
ten dimensions  (or $SL(2,Z)\times R$, in nine
dimensions).  This is manifest in Einstein frame whereas the
preceding expressions were written in string frame.  The
transformation to Einstein frame requires the replacement of the
ten-dimensional string metric, $g_{\mu \nu}$ by $e^{\phi_B/2}\, g_{\mu\nu} =
\Omega_2^{-1/2}\, g_{\mu\nu}$,
which implies that $s$ is  replaced  by $\Omega_2^{1/2}\, s$.  After
taking into account the fact that $A_4$ contains a density factor
$\sqrt{- g}$ we find that (\ref{schemaf}) and (\ref{twothresh}) become, in
Einstein frame,
\ba
A^{E\, an}_4 &\sim&  {\alpha'}^{-4}\, A_4^{Born} +\left( {\alpha'}^{-1} \, Z_{\threeh} +
 {\alpha'} \, Z_{\fiveh}\, s^2
+ {\alpha'}^2 \, \calE_{(\threeh,\threeh)}\, s^3 + {\alpha'}^3\, Z_{\threeh}\,
s^4 \ln(\Omega_2^{\half}) \right. \nn\\
&&\left. + {\alpha'}^4\,  s^5
\ln^2(\Omega_2^{\half})+\cdots\right)\,
\hR^4\, .
\label{schemafe}
\ea
and
\ba
A_4^{E\, nonan} &\sim& \left(s\ln(-\alpha' s) + {\alpha'}^3\, Z_{\threeh}\,
s^4 \ln(-\alpha' s) + {\alpha'}^4\,  s^5
\ln^2(-\alpha' s)+ {\alpha'}^4\,  s^5 \ln(-\alpha' s) \ln(\Omega_2^{\half})\right.
 \nn\\
&&
\left.+
     {\alpha'}^5\, Z_{\fiveh}\,
s^6 \ln(-\alpha' s)+ {\alpha'}^6\,
(Z_{\threeh}\times Z_{\threeh} + \calE_{(\threeh,\threeh)}) \,  s^7
\ln(-\alpha' s)
 +\dots\right)\,
\hR^4\, .
\nn\\
\label{thresheinst}
\ea
 All the factors involving explicit $\Omega_2$'s have disappeared, but
 additional terms with factors of $\ln \Omega_2$ are now present.
 These contribute to the $\ln \Omega_2$ terms that are analytic in the Mandelstam
 invariants in $A^{E\, an}_4$.  The above equations are schematic and omit
 the dependence on $t$ and $u$.  Inserting this, and using $s+t+u=0$, accounts for the
 absence of the $s\ln \Omega_2$ term in (\ref{schemafe}).

Certain terms in (\ref{schemafe}) and (\ref{thresheinst}) are manifestly $SL(2,Z)$ invariant,
but those containing factors of $\ln \Omega_2$  do not transform properly since
 $\ln \Omega_2 \to \ln(\Omega_2/|c\Omega +d|^2)$, under the
$SL(2,Z)$ transformation,
$\Omega \to (a\Omega + b)/(c\Omega + d)$,
($a,b,c,d \in Z$ with $ad-bc=1$).
The lowest order term
of this kind is ${\alpha'}^3\, Z_{3/2}\,s^4 \ln(\Omega_2^{1/2})$, which translates into
a $D^8 R^4$ interaction.  Modular invariance
of the amplitude obviously requires
this to be part of a modular function, which remains to be determined.
We expect that this modular function should be the
solution of a Poisson equation on moduli space, generalizing
the structure of the $D^6 R^4$ interaction (as will be discussed further in
\cite{grv:Poisson}).

\subsubsection{Terms that sum to the ten-dimensional $s \ln (-\alpha' s)$ threshold}
\begin{figure}[h]
\centering
\includegraphics[width=10cm]{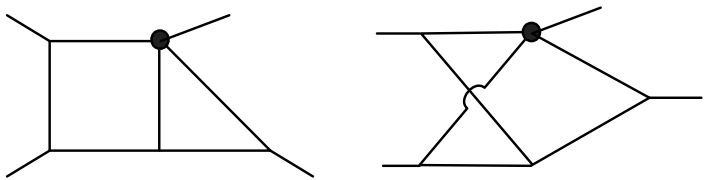}
\caption{Two diagrams with one-loop subdivergences that contain terms that contribute
to the $s \ln (-\al s)$ threshold in the ten-dimensional IIB limit.}
\label{fig:D6R4}
\end{figure}

The leading two-particle threshold, which  arises at genus $h=1$ is just the
massless supergravity normal threshold, which has the form $(-\alpha' s)^{1/2}$ in nine dimensions
and $s\ln(-\alpha' s)$ in ten dimensions.  The change in analytic behaviour is due
to a condensate of the massive Kaluza--Klein
two-particle thresholds that can be symbolically represented
by $\sum_n c_n (-s+ 4n^2/r_B^2)^{1/2}$.  For finite $r_B$ this has a derivative expansion
expressed as the sum of terms of the form $\sum_{k=2}^\infty d_k r_B^{-1+2k}\, D^{2k} \hR^4$
whereas as $r_B\to \infty$ the sum
over $n$ generates the $s\ln(-\alpha' s)$ behaviour of the ten-dimensional theory.  The
analogous infinite
series of dilaton-independent
powers of $sr_A^2$ is explicit in the second line of equation~(3.29) in
\cite{gkv:twoloop}, and will be reviewed in subsection~\ref{iiathreshone} as well as in
appendix~\ref{app:thresh}.
In the IIB case  these terms arise from supergravity loops with different values
of $L$ (whereas, as we will see, in the IIA case the whole series is
contained purely in the one-loop ($L=1$) supergravity amplitude).  The power of $r_B^{-1+2k}$
is obtained by setting  $\q =2k$ in  (\ref{twobact}).  Furthermore, taking
$f_{(\q ,k)}$ to be constant ensures the  genus one condition.
 The expression for these terms in eleven-dimensional coordinates is proportional to
\be
\int d^9 x\sqrt{-G^{(9)}}\,\calV^{1-2k}\, D^{2k} \hR^4\,,
\label{elevmult}
\ee
where $k>1$.  The supergravity  diagrams
that contribute to these terms are those for which
\be
3L = 2k+2 +\frac{m}{3}\, .
\label{lvals}
\ee
Since $L$ is an integer, equation (\ref{lvals})
only has solutions for values of $m$ that are multiples of 3.
We see that the terms in this series have $L>2k/3$.
The lowest value arises at two loops ($L=2$) for $k=2$ and $m=0$.
The precise value of this finite
two-loop diagram was evaluated in \cite{gkv:twoloop} and gave exactly the same result as the
$k=2$ term in the IIA theory expression (\ref{powers}), with $r_A$ replaced by $r_B$.  This
agreement suggests that the higher-loop contributions, which are obtained by increasing
$m$ by multiples of $9$, do not contribute to the $D^4 R^4$ interaction, thereby explaining
why the two-loop result is  exact.  The $D^6R^4$ interaction ($k=3$) arises from $L=3$
with $m=3$, which corresponds to a one-loop subdivergence of a three-loop diagram
as in figure~\ref{fig:D6R4}.

In the IIA case we will see that all the terms that resum to the threshold cut arise from
the one loop ($L=1$) supergravity amplitude.

\subsubsection{Terms that sum  to the ten-dimensional ${\alpha'}^3 \,s^4 \ln (-\alpha' s)$
threshold at genus one and two}
We can also determine in surprising detail properties of
 the infinite series of contributions that sums up to give
 the $s^4 \ln (-\alpha' s)$ threshold required
 by unitarity in the
ten-dimensional IIB theory at genus  one and two ($h=1$ and $h=2$).  These must arise
by summing the Kaluza--Klein thresholds in nine dimensions of the form
$(-s + 4n^2/r_B^2)^{7/2}$.  The low energy expansion is now a series of
terms that
have the form $r_B^{2k-7} \, D^{2k} \hR^4$ in (\ref{twobact}) (again this will be seen  explicitly in the
IIA case in section~\ref{sec:twoadiverge}.
The $r_B$-dependence requires
$\q =2k-9/2$, which results in a dilaton prefactor of $e^{3\phi_B/2}$, which is independent of
$k$.  This means that the modular functions in these terms in (\ref{twobact}) must have
expansion of the form
\be
f_{(2k-{9\over 2},k)}(\Omega,\bar \Omega)= c_0\, \Omega_2^{\threeh} + c_1 \Omega_2^{-\half} + non-pert\, .
\label{modexserII}
\ee
There can be no higher powers of $\Omega_2^{-2}$ since these would generate $s^4\ln (-\alpha' s)$
threshold behaviour at higher genus, which is not permitted by unitarity.
 So we conclude that for all $\q $, $f_{(\q ,k)}$ must be a $SL(2,Z)$ scalar function that has an
 expansion of the form (\ref{modexserII}).  But this is completely consistent with the
 identification
 \be
 f_{(\q ,k)}(\Omega,\bar\Omega) = c_{\q ,k}\,  Z_{\threeh}(\Omega,\bar\Omega)\, ,
 \label{fsdefb}
 \ee
 for some constant $c_{\q ,k}$ that we can fix by requiring the sum of the infinite series to
 reproduce the $O({\alpha'}^3)$ term in the expansion of the
  ten-dimensional unitarity expression (\ref{unigen}).
  When expressed in supergravity coordinates, the terms that contribute to this threshold
  have the form
  \be
  \label{secondthresh}
\int d^9 x\sqrt{-G^{(9)}}\,\calV^{1-2k+{9\over 2}}\,Z_{\threeh}(\Omega,\bar\Omega)\,  D^{2k} \hR^4\, .
  \ee
In this case the number of string loops is related to $k$ by
\be
3L = 2k-4 + \frac{m}{3}\, .
\label{lrelaiib}
\ee
For example, there is a finite ($m=0$)  contribution from two loops ($L=2$)
to $D^{10}R^4$, which has $\q =2k- 9/2=11/2$.

In order to obtain the  $s^5 \ln^2 (-\alpha' s)$ threshold  at genus-two ($h=2$) required
 by unitarity we need to sum a series of terms with powers $r_{B}^{2k - 9}$, which requires
 $\q  = 2k-6$.  This leads to a dilaton factor of $e^{2\phi_B}$ in (\ref{twobact}).
 Therefore, if $f_{(\q ,k)}$ are constants in this case, for all values of $\q $ and $k$ the sum will
 contribute only at genus two, as required.  However, we have not studied the set of coefficients
 that generate a $\ln^2(-\alpha' s)$ factor.

In order to obtain the  ${\alpha'}^5 s^6 \ln  (-\alpha' s)$ threshold  required
by unitarity we need to sum a series with powers $r_{B}^{2k - 11} D^{2k}\hR^4$ in (\ref{twobact}).
The $r_B$-dependence requires
$\q =2k-15/2$, which results in a dilaton prefactor of $e^{5\phi_B/2}$, which is independent of
$k$.  This means that the modular functions in these terms in (\ref{twobact}) must have
expansion of the form
\be
f_{(2k-{15\over 2},k)}(\Omega,\bar \Omega)= c_0\, \Omega_2^{\fiveh} + c_1 \Omega_2^{-\threeh} + non-pert\, .
\label{modexserI}
\ee
There can be no higher powers of $\Omega_2^{-2}$ since these would generate $s^6\ln (-\alpha' s)$
threshold behaviour at higher genus, which is not permitted by unitarity (as can easily be seen by
substituting (\ref{schemaf}) into (\ref{unigen})).
 So we conclude that for all $\q $, $f_{(2k-15/2,k)}$ must be a $SL(2,Z)$ scalar function that has an
 expansion of the form (\ref{modexserI}).  But this is completely consistent with the
 identification
 \be
 f_{(2k-{15\over 2},k)}(\Omega,\bar\Omega) = c\,  Z_{\fiveh}(\Omega,\bar\Omega)\, ,
 \label{fsdefbII}
 \ee
 for some constant $c$ that can be fixed by requiring the sum of the infinite series to
 reproduce the $O({\alpha'}^3)$ term in the expansion of the
  ten-dimensional unitarity expression (\ref{unigen}).
  At higher orders in $\alpha'$ than those considered up to this point the analysis
  rapidly gets very complicated.

\section{Higher derivative interactions in type IIA}
\label{sec:twoa}

Whereas the IIB coupling is the dimensionless ratio of the radii of the torus,
the IIA coupling is determined by the single length scale $R_{11}$. This will lead to
powerful restrictions on the possible powers of the coupling
in the IIA action (\ref{twoaacttwo}).

\subsection{Terms that are finite as $r_A\to \infty$ -- non-renormalisation conditions}
\label{sec:twoafin}

To be specific we
will first specialize to those terms that are proportional to $r_A$ and so
have a finite ten-dimensional ($r_A\to \infty$)
 limit.  Therefore
$\q $ will be taken to satisfy (\ref{largera}),
\be
2\q =3+k-3h\, ,
\label{onesdef}
\ee
or, from the definitions (\ref{ksdef}),
\be
\w = 3+3k -3h -2\beta_L\, .
\label{boundks}
\ee
Since $\w\ge 0$ we see that
\be
h \le k  +1- \frac{2\beta_L}{3}\, .
\label{hbound}
\ee
The ten-dimensional limit for a term with string genus $h$ has the form
\ba
S^{(k+3)A}_{m,h}
 = l_s^{2k-2} \, \hLambda^{m} \int d^{10} x \sqrt{-g_A}\,
c_h\, e^{2(h-1)\phi_A}\,   D^{2k}\hR^4\,
\label{twoaactten}
\ea
(recall that $m =9L-6-2k-2q = 3(3L-3-k+h)$).
Since $\beta_1=0$ while $\beta_L\ge 2$ for $L\ge 2$ 
we need to distinguish the case $L=1$ from $L>1$.

(a) When  $L=1$ we need to choose $\beta_L=0$ in (\ref{boundks}) so that $h\le k +1$.
If $h=k+1$ it follows that $\w=0$ and so $m=3$ and  (\ref{minkelev}) contains
a factor of $\Lambda^3$, corresponding to the one-loop divergence.
However, the $\Lambda^3$ divergence of the one-loop amplitude
\cite{Russo:1997mk,gkv:twoloop} is independent of the momenta and only contributes to the
$k=0$, $h=1$ term.
The $h=k$ terms have $\w=3$ and are finite ($m=0$ so
(\ref{minkelev}) is independent of $\Lambda$).
The contributions of these terms, given in \cite{Russo:1997mk,gkv:twoloop}, is
\begin{equation}
\label{genuses}
2\zeta(3) R^4+8\pi^2\sum_{h=2}^\infty l_s^{2h-2}\, \int d^9 x \sqrt{-g_A^{(9)}}\, r_A\, \Gamma(h-1)
 \zeta(2h-2) {e^{2(h-1)\phi^A}\over h!} D^{2h} \, \hR^4 \,.
\end{equation}

(b) For $L>1$ we are assuming  $\beta_{L}\ge 2$  and so it follows  from
(\ref{hbound}) that
\be
h < k -\frac{1}{3}\, .
\label{hkless}
\ee

Clearly, these ten-dimensional IIA results could have been obtained more directly by
considering compactification from eleven to ten dimensions
on a circle of radius $R_{11}$, instead of considering a torus. In that case
the starting expression (\ref{twotorus}) is replaced by
 \begin{equation}
S^{(\beta_L+\v+3)}_L  =
l_P^{9(L-1)} \, \sum_{\w=0}^{\w_L}\Lambda^{9L-6-2\beta_{L}-w} \int d^{10} x \sqrt{-G^{(10)}}\,
 R_{11}^{1-\w } \,(R^2_{11} D^2)^\v\, D^{2\beta_{L}}\hR^4\ .
\label{circlecom}\end{equation}
Translating into IIA string variables immediately leads to~(\ref{twoaactten})
with the same set of equations~(\ref{onesdef}) -- (\ref{hbound}) (where
$k=\v+\beta_L$ and $\q  = \w/2 -v$, as given in~(\ref{ksdef})).

We therefore conclude that type IIA string theory in ten dimensions satisfies the following strong
conditions (for $k\geq 1$):
\begin{itemize}
\item {\it There are no contributions with string loop genus $h >k$}.

\item  {\it The contributions with $h=k$ are determined exactly by finite contributions
from the derivative expansion of the one-loop ($L=1$) diagram in eleven dimensions}.

\item  {\it Contributions with $h<k$ are permitted and may arise from any number of
supergravity loops greater than one ($L>1$)}.

\end{itemize}

It is of interest to consider what would happen if extra powers of $D^2\sim S,T,U$ were to factor out
of the $L$-loop amplitudes of eleven-dimensional supergravity for $L>2$ (in other words, if $\beta_L>2$
for $L>2$).  For example,
if it turned out that every extra loop had an extra power of $D^2$ then we would have
$\beta_L=L$.  In that case (\ref{boundks}) would imply $\w = 3+3k-3h-2L$, so that $h \le k+1 -2L/3$.
This would mean that the bound still requires $h\le k $.  Furthermore,
since $h\ge 0$,  all perturbative contributions
to $D^{2k}R^4$ would be obtained  from $L \le  3(k+1)/2$.

Although we have obtained an upper bound on $h$, we have not shown that it also satisfies
the obvious lower bound, $h\ge 0$, that ensures there are no terms more singular than the genus-zero tree-level terms.
Nevertheless, there are indications from the explicit $L=1$
and $L=2$ calculations that contributions with $h<0$ are, in fact, absent.
Note also that the type IIA bound does not explain why certain terms are absent in string
perturbation theory, such as
$D^4 R^4$ at one loop ($k=2$, $h=1$).  In the IIB case, where the perturbative contributions to  $D^4R^4$
 are contained in $Z_{5/2}$, the  $h=1$ component is explicitly absent.

Since the considerations up to this point have
mostly been purely dimensional one might expect similar statements to hold for
other terms of the same dimension.  For example, this ties in with the
argument in \cite{Russo:1997mk} that powers of the curvature of the form $R^{3n+1}$ are
the only ones that have a nonzero eleven-dimensional limit.
 Although the case
$k=1$  is absent for $D^2R^4$ since  $s+t+u=0$, a term of the same dimension might be present
 and if the above arguments extend to this case, it would have
a contribution of maximum possible genus $h= 1$.

One consequence of our statements is that  the  low energy behaviour of the genus-$h$
loop contribution to the type IIA
 four-graviton effective amplitude has the form $s^h \hR^4 (1 +O(s))$ (when $h>1$).
This is a very powerful condition, which indicates that the low-energy
  ten-dimensional theory is much less divergent than it might have
  been.  This will be discussed in more detail in section~\ref{sec:discuss}.

\subsection{Terms that diverge as $r_A\to \infty$}
\label{sec:twoadiverge}

A general nine-dimensional contribution behaves as $r_A^{1+p}$, which implies the relation
(\ref{largera}) and leads to a generalization of the bound (\ref{hkless}) of the form,
$h\le k -1/3 -p/2$.  Terms that diverge in the $r_A\to \infty$ limit have $p>0$ so the bound is
stronger.
Just as for the type IIB theory, these are the terms that arise from the low energy expansion of
 the normal thresholds due to massive Kaluza--Klein
intermediate states.    These terms must sum up to give, in the $r_A\to \infty$ limit,
the appropriate ten-dimensional threshold behaviour.
The unitarity relation  for the ten-dimensional
IIA case has identical perturbative structure to the IIB case (at least to the order considered
explicitly in the IIB case) so it is given by (\ref{unigen})
with the omission of the non-perturbative $D$-instanton contributions.

\subsubsection{Terms that sum to the ten-dimensional $s \ln (-\alpha' s)$ threshold}
\label{iiathreshone}
Just as in the IIB case, the IIA theory must generate
a non-analytic $ s\ln (-\alpha's)$ genus-zero massless supergravity
threshold in the four-graviton amplitude
in ten dimensions. Once again, the leading
massless supergravity normal threshold has the form $(-\alpha' s)^{1/2}$ in nine dimensions
and $s\ln(-\alpha' s)$ in ten dimensions.  The infinite
series of dilaton-independent
powers of $r_A^{-1+2k}$ is explicit in the second line of equation~(3.29) of
\cite{gkv:twoloop} and has the form
\begin{eqnarray}
r_{A}^{-1} (-\al s)^{1\over2}\, \hR^4&-&\pi^{-{1\over2}}\,
\sum_{k=2}^\infty
  {\Gamma(k-\half)\over k!}\,  \zeta(2k-1)\,r_A^{2(k-1)}\,(-\al s)^{k}\hR^4  \, .
\label{powers}
\end{eqnarray}
These contributions arise from the expansion of the one-loop amplitude, and resum to the
 $s\,\ln(-\alpha' s)$ threshold contribution in the
limit $r_{A}\to\infty$ (as  reviewed  in the Appendix~\ref{app:thresh}).
In order to reproduce this in (\ref{twoaactone}) we need
\be
r_A^{1 - \frac{4\q }{3} + \frac{2k}{3}} = r_A^{2k-1}\, ,
\label{genzterms}
\ee
or $2(k+\q )=3$, which is the same condition as for the series of $h=k$ terms
leading to (\ref{genuses}).
The two sets of terms (\ref{genuses}) and (\ref{powers}) are very similar when expressed in
eleven-dimensional coordinates -- they
differ only by a `9-11' flip
that interchanges $R_9$ and $R_{11}$, thereby interchanging the roles of $r_A$ and $e^{\phi_A}$.

\subsubsection{Terms that sum  to the  ten-dimensional ${\alpha'}^3 \,s^4 \ln (-\alpha' s)$
threshold at genus one and two}

As in the IIB case, we here want to sum a series of terms of the form $r_A^{2k-7}D^{2k}\hR^4$.
There should be a contribution with $h=1$ and one with $h=2$.
When $h=1$ we see from
(\ref{twoaacttwo}) that we need to set $1-4\q/3+2k/3 = -1+2k-6$ in order to get the correct power
of $r_A$.
This means that $\q +k = 6$, so that there is contribution from two supergravity loops ($L=2$) with $m=0$,
since $m=9L-6-2(k+q)$.
The $h=2$ term is obtained by setting $\q +k = 9/2$ in order to get the correct
power of $r_A$.  In this
case the value of $L$ can only be an integer if $m=3\, {\rm mod}\, 9$.
Setting $m=3$ again results in $L=2$, but now the contribution comes from a $\Lambda^3$ subdivergence, which
agrees with the fact that the one-loop
subdivergences of the two-loop
diagrams generate the correct series of terms to reproduce the
ten-dimensional $h=2$ threshold.

\begin{figure}[h]
\centering
\includegraphics[width=15cm]{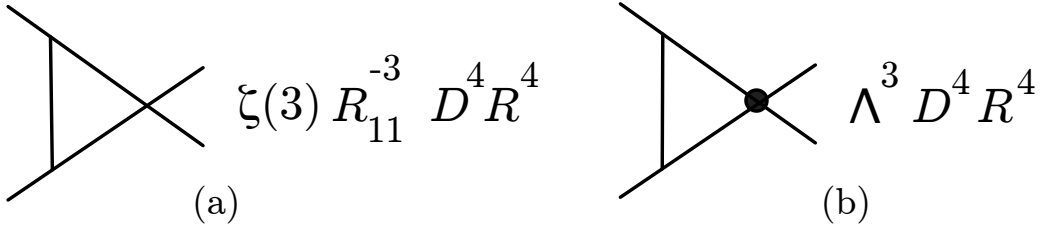}
\caption{The string one-loop $s^4\ln(s)$ originates from the two-loop supergravity amplitude by
resumming the Kaluza-Klein modes
in one loop and using the finite $\zeta(3) \, D^4 \hR^4/R_{11}^3$ contribution from the one-loop amplitude, as in figure (a).
The string two-loop contribution
arises by picking the $\Lambda^3$ counterterm at one-loop, as in figure (b).}
\label{fig:s4logs}
\end{figure}

Figure~\ref{fig:s4logs} depicts the origin of such threshold terms from two-loop supergravity Feynman diagrams.
The string genus-one ($h=1$) $s^4 \ln(-\al s)$ threshold behaviour comes from resumming the
Kaluza-Klein states in one loop while summing over non-zero windings in the second (this picks out the finite part of
the second loop proportional to $R_{11}^{-3}$).
The string genus-two ($h=2$) part of the  $s^4 \ln(-\al s)$ threshold
is obtained by taking the zero winding number sector (proportional to the $\Lambda^3$ subdivergence) in the second loop,
and also including  the diagram in which a
counterterm replaces that loop.  The contributions in the figure are therefore proportional to each other, with coefficients
that are determined by the coefficients of the tree-level and one-loop terms in the $R^4$ interaction that are
contained in $Z_{3/2}$.

Both the diagrams on the right side of figure~\ref{fig:s4logs} are proportional to
\begin{equation}
 \left[S^2 I_{Triang}(S)+ T^2  I_{Triang}(T)+
U^2  I_{Triang}(U)\right] \, \hR^4\, ,
\end{equation}
where $I_{Triang}(S)$ is a scalar field theory triangle diagram.
In appendix~\ref{app:thresh} we demonstrate that in ten dimensions this triangle diagram has the  non-analytic
structure, $I_{Triang}(s)\sim s^2 \, \ln(-\al s)$.
When multiplied by the appropriate $h = 1$ factor, $\zeta(3)\, e^{-2\phi_A}\, s^2\, \hR^4$
(from the finite part of the $L=1$ one-loop amplitude), or the $h = 2$
factor, $\zeta(2)\, s^2\, \hR^4$ (from the regulated divergence of the $L=1$ one-loop amplitude),
this gives the expected string threshold contributions.
 The $h=1$ piece reproduces the $s^4 \ln (-\al s)$
term found by explicitly analyzing the one-loop string amplitude in  appendix~B of \cite{gv:stringloop}.

On the other hand in nine dimensions with finite $r_A$ we see from
(\ref{e:triangleexp})   that  the triangle diagram
can be expanded in the expected infinite series of positive powers of $r_A$ that again sum up to
the nine-dimensional thresholds for Kaluza--Klein states,
\begin{eqnarray}
\nn  (-\al  s )^{3\over2} &-&
\sum_{k=2}^\infty\,  {\Gamma(k-\threeh) \over \Gamma(-\threeh)\, 2^{2k-3}\, k!}\zeta(2k-3)
  r_A^{2k-3}\,( \al  s )^{k}\\
  & =& \sum_{n\in\ZZ}
  \left( \frac{4n^2}{r_A^2}-\al s \right)^{3\over2}\, .
\label{powersapp}
\end{eqnarray}
Multiplying this contribution by $\zeta(3)\, s^2 \hR^4$ gives the $h=1$ threshold term,
\begin{eqnarray}
\lim_{r_{A}\to\infty} \zeta(3)\, r^{-1}_{A}\, \sum_{n\in\ZZ}
  \left( \frac{4n^2}{r_A^2}-\al s\right)^{3\over2}\, s^2 \hR^4
  \sim 2^{-5}\, \zeta(3)\, (\alpha'\,s)^4\,\ln(-\alpha'\,s)\, \hR^4\,
\label{onethreshh}
\end{eqnarray}
(where we have only kept the $\ln (-\al s)$ term on the right-hand side).
Similarly, the $h=2$ part is simply reproduced by multiplying by the regulated $L=1$ one-loop
divergence
 $\hLambda^3 \, s^2\hR^4\sim\zeta(2)\, s^2 \hR^4$.
%

\subsection{Lifting to eleven dimensions}\label{sec:Mtheory}

Now that we have seen how the string theory expansion is constrained by duality with eleven-dimensional
supergravity we can ask how self-consistency restricts the eleven-dimensional theory -- thereby
completing the circle.  In other words, which higher derivative interactions survive decompactification
to eleven dimensions?  This is answered by reexpressing
the ten-dimensional IIA terms in (\ref{twoaactten}) in terms of eleven-dimensional
supergravity on a circle as
 \be
S_h^{(k+3)} =l_P^{9L-6-3h+k}
\Lambda^{9L-6-2k}\,\int d^{10}x\, R_{11}\sqrt{-G^{(9)}}\,R_{11}^{3h-3-k}\, D^{2k}\hR^4\,.
\label{elevenfroma}
\ee
The terms that have finite non-zero large-$R_{11}$ limits are ones for which
\be
\label{khcond}
k=3h-3\, .
\ee
All other terms either vanish or diverge as $R_{11}\to \infty$.  As we saw when passing from nine to ten
dimensions (and as discussed in \cite{gkv:twoloop}) the sum of divergent terms must generate the
non-analytic thresholds in eleven dimensions that contribute to nonlocal parts of the action.

We now list some examples of terms satisfying (\ref{khcond})
that contribute to local terms in the eleven-dimensional action
and how they emerge from the supergravity calculations.
\begin{itemize}
\item
The $R^4$ interaction ($k=0$) only gets a contribution from  one loop
($L=1$) eleven-dimensional supergravity
(since all $L>1$ have $k> 0$).  This is the genus-one ($h=1$) term that has
$m=3$ and so is proportional to $\Lambda^3$.  The precise value of the counterterm
that cancels this divergence was determined in \cite{ggv:oneloop}.
\item
The first non-zero value of $k$ satisfying (\ref{khcond}) is $k=3$, which contributes to the
genus-two ($h=2$) IIA string action.  From  (\ref{twoaactten}) we see that this gets a
contribution from two-loop supergravity ($L=2$) with $m=9L-6-2k=6$, so it is
proportional to $\Lambda^6$.  The precise value of the counterterm needed to subtract this dependence
on the cut-off was determined in \cite{gv:D6R4}, where the corresponding $h=2$ coefficient of the
$D^6R^4$ interaction was determined in the IIB theory.  The known equality of the IIA and
IIB four-graviton string theory amplitudes at two loops therefore fixes the coefficient in the
IIA theory, which then lifts to eleven dimensions.   As mentioned earlier, the precise matching
of the other $D^6R^4$ coefficients in the IIB theory encourages us to think that this value is
not renormalised by higher loops ($L>2$) although we do not have an explanation for this.
  \item
The next possible value of $k$ is $k=6$.  From (\ref{khcond}) this corresponds to a genus-three ($h=3$)
string theory term.  This is a logarithmically divergent ($m=0$) contribution to the two-loop
($L=2$) $D^{12}R^4$  term
in (\ref{twoaactten}).  In this case
we know of no reason why the value of this coefficient should be protected against contributions from
 higher values of $L$.
\end{itemize}

A plausible generalization of this argument would
suggest that only terms of the form $D^{6h-6-2n}R^{4+n}$ and other terms of the same
dimension can appear in the M-theory
effective action after taking the eleven-dimensional limit
$R_{11}\to\infty $, as was suggested in \cite{Russo:1997mk}.

\section{Reduction to lower dimensions.}
\label{sec:discuss}

In the preceding sections we studied general features of $L$-loop Feynman diagrams of
four-graviton scattering in eleven-dimensional supergravity
compactified on a two-torus and the consequences, via duality, for type IIA and IIB string theory
compactified on a circle to nine dimensions.  The Feynman diagrams beyond $L=2$ are
extremely complicated and their detailed structure was not used in our discussion, other than the fact
the leading term in the low energy expansion behaves as $D^{2\beta_L} R^4$, where $\beta_1=0$, $\beta_2=2$
and $\beta_L \ge 2$ for $L>2$.  Obviously, since eleven-dimensional supergravity is not renormalisable it does
not, by itself, give a well-defined quantum theory.  In particular, short-distance properties
smuggled into the momentum cut-off can only be determined with additional input.  In our
discussion the input consisted of the requirement that the toroidally compactified theory
should be equivalent to string theory via the usual duality considerations.  As a consequence,
we found a number of interesting constraints on both the IIA and IIB theories, as well as
consistency conditions on higher derivative terms in the eleven-dimensional theory itself.

Among these constraints was a strong non-renormalisation condition on the type IIA
derivative expansion. In section~\ref{sec:twoafin} we found that the ten-dimensional II string perturbation
theory amplitude at genus $h$ has a low energy limit that begins with a term of the form
$D^{2h} R^4$.  Gratifyingly, this has also been verified, at least up to $h=5$, directly in string
perturbation theory in \cite{Berkovits:2006vc} (where the $D^{2k} R^4$  terms with $k<6$ are viewed as
analogues of `F-terms' within the Berkovits formalism).  This means that type IIA supergravity -- the low energy
limit of ten-dimensional IIA string theory -- is more finite than might have been expected.
In order for the leading $h$-loop behaviour to be $D^{2h}\, R^4$  there must be cancelations between the many
Feynman diagrams that contribute to the $h$-loop amplitude of
ten-dimensional IIA supergravity that extend the cancellations  that arise at one and two loops and result in an extra
factor of $D^2$ for every additional loop.

More explicitly, the Feynman diagram expansion of ten-dimensional IIA supergravity can be organized as a loop
expansion in powers of the string coupling, $e^{\phi_A}$, and so we have the leading
contribution, $S^{(3+h)}_{(10)}$  at any genus $h$ of the form (for $h> 1$)
\begin{eqnarray}\label{tendim}
S^{(3+h)}_{(10)}&=& l_s^{8h-8}\, \Lambda_{10}^{6h-6}\, \int d^{10}x \, e^{2(h-1)\phi_A}\,  \sqrt{-g_A}\, D^{2h} \hR^4\\
\nn&=& \kappa_{10}^{2(h-1)} \, \Lambda_{10}^{6h-6}\, \int d^{10}x \,  \sqrt{-g_A}\, D^{2h} \hR^4
\end{eqnarray}
 which
 is analogous to the eleven-dimensional $L$-loop term in (\ref{minkelev}) but with $L\to h$,
 $\beta_L \to h$  and a
 cut-off $\Lambda_{10}$ (and $w=0$).
 The second line gives the expression as an expansion in the ten-dimensional Newton constant $\kappa_{10}^2 = l_{10}^8$,
 where the ten-dimensional Planck scale is related to the string scale by $l_{10}= e^{\phi_A/4}\, l_s$.
   Although the counterterms that cancel the divergences in this expression are undetermined
   the first line has the same form as
 the precise expression for the $h$-loop contribution to the $D^{2h}R^4$ term of ten-dimensional IIA string
theory given by (\ref{genuses}).

In order to understand the connection with our starting point in eleven dimensions
we can refer (\ref{tendim}) back to our earlier M-theory units  using $R_{11}=l_{s} \exp(\phi_{A})$
and $l_{P}=l_{s}\, g_{s}^{1/3}$, which gives
\begin{equation}\label{kkoneloop}
S^{(3+h)}_{(11)}=  {1\over R_{11}^3}\,(l_{P}\Lambda_{10})^{6(h-1)}\,
(l_{P}R_{11}^{-1})^{3(h-1)}\, \int d^{11}x\, \sqrt{-G}\, (R_{11}^2 D^2)^h \hR^4\, .
\end{equation}
If we relate the eleven-dimensional and ten-dimensional cut-offs by  $\Lambda^3=\Lambda_{10}^2/R_{11}$,
we see that (\ref{kkoneloop}) has the form of  the compactification of the finite part of the
one-loop ($L=1$) amplitude of the
eleven-dimensional theory on a circle, expanded in powers of $S$, $T$ and $U$
(equation (\ref{circlecom}) with $L=1$, $\beta_L=0$, $\w=3$ and $\v=h$ and a particular choice of counterterm).
This shows that the particular relation between the string parameters and
 the M-theory parameters makes  the perturbation expansion in eleven dimensions look
  very different from the one in lower dimensions.

Now consider the IIA supergravity after dimensional reduction to $d<10$  dimensions.  This can be
implemented, for example, by compactifying the genus-$h$ string theory
amplitude (with its overall factor of $s^h$)
 on a $(10-d)$-torus of scale $r$
and taking the low energy limit $r\to 0$ with $\alpha'/r \to 0$
(with the momenta and polarizations of the external gravitons in the non-compact directions).
Then, for $h>1$, simple dimensional analysis shows that
(\ref{tendim}) becomes
\begin{eqnarray}\label{ddim}
S^{(3+h)}_{(d)}&=& l_s^{(d-2)(h-1)}\, \Lambda_d^{(d-4)h-6}\,
\int d^dx \, e^{2(h-1)\phi_A^{(d)}}\,  \sqrt{-g_A^{(d)}} D^{2h}\hR^4\\
\nn & =&\kappa_{(d)}^{2(h-1)}\, \Lambda_d^{(d-4)h-6}\, \int d^dx \, \sqrt{-g_{A}^{(d)}}\, D^{2h}\hR^4\, .
\end{eqnarray}
 where  $\phi^{(d)}_{A}$ is the $d$-dimensional dilaton and $\Lambda_d$ the
 cut-off parameter of  $d$-dimensional supergravity.
  It should be emphasized that (\ref{ddim})
is a schematic representation of the low energy limit of the four-graviton
amplitude that is relevant when the power of $\Lambda_d$ is positive, indicating the presence of
ultraviolet divergences.   In obtaining (\ref{ddim}) we have assumed that inverse powers of $D^2$ (i.e.,
powers of $1/s$, $1/t$ or $1/u$ in the amplitude) do not arise in the the process of taking the
low-energy limit of the toroidally compactified string amplitude, so that the  power of
$\Lambda_d$ is not increased.  With this assumption, we see  from (\ref{ddim}) that ultraviolet
divergences are absent in dimensions satisfying
\be
d < 4 +\frac{6}{h}\,,
\label{dbound}
\ee
for $h>1$ (while $d<8$ for $h=1$)\footnote{When the bound (\ref{dbound}) is satisfied, the negative
power of $\Lambda_d$
in (\ref{ddim}) is replaced by an expression involving inverse  powers
of $s$, $t$, or $u$, together with logarithms, associated with
infrared effects and leading to infrared divergences
when $d\le 4$.  This structure of the low energy limit of string
theory loop amplitudes
is explicitly illustrated at one loop ($h=1$)  in \cite{Green:1982sw}
and at two loops in
\cite{Bern:1998ug,Smirnov:1999gc,Tausk:1999vh,Anastasiou:2000kp}.}.
 If this bound is indeed satisfied it means that ultraviolet divergences
are absent to all orders in four or less dimensions\footnote{
In the earliest version of this paper we inadvertently stated that (5.3)
might allow logarithmic ultraviolet divergences when d=4, which it manifestly does not.}, so
it seems possible that four-dimensional $N=8$ supergravity is ultraviolet finite.

It is interesting that the
condition (\ref{dbound}) is precisely the same condition as for maximally extended supersymmetric
Yang--Mills theory ($N=4$ Yang--Mills in $d=4$ dimensions)
given in  \cite{Bern:1998ug} (and reproduced in a superspace formulation
 in \cite{Howe:2002ui}).  This is quite
remarkable since the pattern of potential divergences in the two theories is very different.  To begin
with,  pure four-dimensional Yang--Mills theory is renormalisable by power counting in four dimensions.
In the maximally supersymmetric extension the low-energy one-loop four-gluon amplitude is proportional to $F^4$,
while all higher loops are proportional to $D^2 F^4$ with no
extra powers of $D2$ beyond two loops. Given these facts one can easily obtain the
divergence bound (\ref{dbound}).  This striking similarity between super Yang--Mills and
supergravity is presumably connected to the relation between open-string and closed-string theory known since the earliest
days of string theory.  This was exploited a long time ago \cite{Kawai:1985xq}
in order to obtain closed-string tree amplitudes from open-string tree amplitudes in an efficient manner,
and subsequently \cite{Bern:1998ug} has proved  to be of great interest in deciphering the structure of
supergravity perturbation theory beyond tree level.  A number of further
fascinating correspondences between
the structure of  $N=8$ supergravity amplitudes and the amplitudes of  $N=4$ super-Yang-Mills
have since been discovered \cite{Bern:2005bb,Bjerrum-Bohr:2005xx,Bjerrum-Bohr:2006yw, Bern:2006kd} which suggest that $N=8$
might be less ultraviolet divergent than expected.

 However, we should emphasize an important point which was overlooked
in earlier versions of this paper.  In string theory compactified on a torus
there are not only `perturbative' string states
but also `non-perturbative'  $Dp$-branes and  Neveu--Schwarz branes wrapped on the torus as well
as  Kaluza-Klein charges and Kaluza--Klein monopoles.
The low energy limit that is relevant for describing the Feynman diagrams of quantum gravity
(in which $\kappa_4$ is held fixed) keeps only the massless states in the perturbative sector.
However, it is straightforward to see that an infinite subset of the non-perturbative states {\it necessarily}
also becomes  massless.
This could have a profound effect on the nature
of the low-energy limit irrespective of whether the perturbative
loops are or are not UV finite\footnote{Similar observations have been made by H. Ooguri and
J.H. Schwarz (private communication).}.

 Although we have only considered the four-graviton amplitude, it is quite
plausible that several
features that we have discussed also hold more generally.  In particular,
since unitarity, along with supersymmetry, should interrelate all
$n$-particle amplitudes it seems likely that the ultra-violet
finiteness properties of $A_4$ will extend to the complete S-matrix.
This is also supported by the fact that the present power-counting analysis
can be extended in a straightforward way to other terms of the same
dimension.

The considerations of this paper have been
based on compactifying eleven-dimensional supergravity on a two-torus, although there should be generalizations
to compactifications on higher-dimensional tori or more complicated compact spaces.  The richer set of
dualities should, in principle,
lead to further consistency constraints on the string theory derivative expansion, but there are a number of
unresolved issues concerning the r\^ole of $M2$-brane (and $M5$-brane) instantons that raise new difficulties for
such generalizations.

To summarize,
our circle of arguments imposed interconnected consistency conditions on eleven-dimensional
supergravity and ten-dimensional string theory in its low energy limit. We should stress that in order to establish
the validity of our use of perturbative approximations to the eleven-dimensional theory
it would be necessary to develop a much better understanding of the full
equivalence between M-theory and string theory.

Nevertheless, it is
striking  that these arguments suggest that the ultraviolet divergences of maximally extended supergravity
could be milder than might have been anticipated and may even be absent in four dimensions.

\acknowledgments
We are particularly
grateful to Nathan Berkovits for several useful interactions in the course
of this work.  We also wish to thank Costas Bachas, Zvi Bern, Lance Dixon, Eliezer Rabinovici and  Savdeep Sethi
for correspondence and conversations.
 We are also grateful to Hirosi Ooguri and John Schwarz for discussions concerning the role of light
$p$-branes in the low energy limit.
P.V. would like to thank the LPTHE of Jussieu for hospitality where part of this work was carried out.
J.R. also acknowledges support by MCYT FPA 2004-04582-C02-01.
This work was partially supported by the RTN contracts
MRTN-CT-2004-503369, MRTN-CT-2004-512194 and MRTN-CT-2004-005104  and by the ANR grant BLAN06-3-137168.


\appendix
\section{Properties of tree-level and one-loop string theory}
\subsection{ Tree-level four graviton scattering in type II string theory}\label{sec:treelevel}

In this appendix we will review the known coefficients of the higher derivative
interactions that are contained in the tree-level and one-loop string perturbation
theory expressions.  Whereas the tree-level terms are easily obtained to all orders in
$\a'$, the one-loop coefficients are only known to low orders.
The amplitude has the form
\cite{GreenSchwarzloop,GreenSchwarzWitten,DhokerPhongRevue},
\be
\label{etreeone}
A^{(2)}_4 =\kappa_{10}^2\,  \hK \, e^{-2\phi}\, T(s,t,u)\, ,
\ee
where $2\kappa_{10}^2= (2\pi)^7  {\alpha'}^4 $ and $T(s,t,u)$
 is given by
\bea
\label{eTree}
T &=& {64\over l_s^6  stu}
{\Gamma(1-{\alpha'\over4} s)\Gamma(1-{\alpha'\over 4} t)\Gamma(1-{\alpha'\over 4} u)
 \over \Gamma(1+ {\alpha'\over 4} s)\Gamma(1 + {\alpha'\over 4} t)\Gamma(1  +
 {\alpha'\over 4} u)} \nn\\
&=& {64 \over l_s^6 stu} \exp\left(\sum_{n=1}^\infty {2 \zeta(2n+1) \over
 2n+1}{{\alpha'}^{2n+1}\over 4^{2n+1}} (s^{2n+1} + t^{2n+1} + u^{2n+1})\right)\ .
\eea
Define
$$
\s_n=\big({\a'\over 4}\big)^n (s^n+t^n+u^n)
$$
It is easy to see that all $\s_n $ can be written in terms of
$\s_2 $ and $\s_3$ as follows \cite{gv:stringloop}
\be
\s_n=n \sum_{2p+3q=n}{(p+q-1)!\over p!q!}
\left({\s_2\over 2}\right)^p \left({\s_3\over 3}\right)^q\ .
\ee
In terms of $\s_2,\ \s_3$, the low energy expansion of the amplitude reads
\be
T ={3\over \sigma_{3}} + A(\s_2,\s_3)\ ,
\ee
with
\bea
A(\s_2,\s_3)&=&\sum_{p,q=0}^\infty  T_{(p,q)}\s_2^p\s_3^q
\non\\
 &=&  2\zeta(3)+
\zeta(5)\s_2+{2\over 3}\zeta(3)^2\s_3+
{1\over 2}\zeta(7)\s_2^2+{2\over 3}\zeta(3)\zeta(5)\s_2\s_3
\non\\
&+&
{1\over 4}\zeta(9)  \s_2^3 +{2\over
  27}\left(2\zeta(3)^3+\zeta(9)\right)\s_3^2+{1\over 6}(2\zeta(3)\zeta(7)+\zeta(5)^2)\s_2^2\s_3
\non\\
&+&
{1\over 8}\zeta(11)\s_2^4+{1\over
  9}\left(2\zeta(3)^2\zeta(5)+\zeta(11)\right)\s_2\s_3^2+\cdots \, .
\label{longaf}
\eea

The number of kinematical structures appearing
at each order $D^{2k}\hR^4$ is given by
 the number of ways $k$
decomposes as the sum of a multiple of 2 and a multiple of 3,
$k=2p + 3q$
(so that $\s_2^p\s_3^q$ corresponds to the order $s^k \hR^4$).

\subsection{One-loop four-graviton scattering in type II string theory}\label{sec:oneloop}

The one-loop four graviton scattering is given by an integral over the complex positions $\nu_i$ ($i=1, 2,3,4$)
of four vertex operators on a toroidal world-sheet with complex structure $\tau$, which is to be integrated
over the fundamental domain \cite{GreenSchwarzloop},
\begin{equation}\label{Afoneloop}
A_{4}^{one-loop} ={\kappa_{10}^4\over 2^5\, \pi^6\, {\alpha'}^4} \, \hR^4\,\int_{{\cal F}}
{d^2\tau\over \tau_{2}^2}\, \int_{\cal T}
 \prod_{i=1}^4{d^2\nu_{i}\over \tau_{2}}\,  (\chi_{12}\chi_{34})^{\alpha's} \,
 (\chi_{14}\chi_{23})^{\alpha't}\, (\chi_{13}\chi_{24})^{\alpha'u} \ ,
\end{equation}
where $\chi_{ij}$ is the scalar Green function between the points $\nu_{i}$ and $\nu_{j}$ on the world-sheet
torus.
The low-energy expansion of this amplitude can be evaluated  order by order in $\alpha'$ using a diagrammatic
method described in \cite{gv:stringloop}.  Technical difficulties arise at order ${\a'}^4 s^4$
 due to the presence of logarithmic massless threshold  singularities.
 This expansion is discussed in detail in
\cite{grv:stringloop}, where the expansion is carried out up to and including order ${\a'}^4 s^4$ giving
\begin{eqnarray}\label{eOneLoop}
A_{4}^{one-loop}& =& {\kappa_{10}^4\over 2^5\, \pi^6\, {\alpha'}^4} \, \hR^4\, {2\zeta(2)\over \pi} \,
\left(1+ 0\cdot\sigma_{2} + {\zeta(3)\over 3}\, \sigma_3
+ 0\cdot \, \sigma_{2}^2
+  O(s^5)\right)\, .
\end{eqnarray}
It is notable that the coefficients of the $D^4\hR^4$ and $D^8\hR^4$ terms vanish.

\section{Ten-dimensional thresholds from nine dimensions}
\label{app:thresh}

In this appendix we provide some details of how the nonanalytic terms of the ten-dimensional
IIA theory arise by resumming
and infinite series of terms in nine dimensions and taking the limit $r_A\to \infty$.
The two examples
discussed in the main text are (i) the one-loop box diagram (in figure~\ref{fig:BoxDiagram})
that gives $s\,\ln(-\al s)$,
and (ii) the triangle diagram
that contains a one-loop counterterm  (in figure~\ref{fig:s4logs}), which contributes to $s^4\, \ln(-\al s)$.

(i) The one-loop $\varphi^3$ scalar field theory box integral compactified on a two-torus from eleven to nine dimensions
is the sum of three terms, $I_{Box}(S,T) + I_{Box}(T,U) +I_{Box}(U,S)$, containing threshold singularities
in ($S,T$), ($T,U$) and ($U,S$), respectively.  The function $I_{Box}(S,T)$ is given by
\cite{Russo:1997mk,gkv:twoloop}
\begin{equation}
I_{Box}(S,T)={2\pi^9\over l_{P}^2\mathcal{V}}\, \int_{0}^\infty {dt\over t^{3\over2}}
\, \int_{\mathcal{T}_{ST}}\, \prod_{r=1}^3 dw_{r}\,
\sum_{m_{I}=(m_1,m_2)}\, e^{- G^{IJ}m_{I}m_{J}\, t- Q(S,T)\, t}
\end{equation}
where $\mathcal{T}_{ST}=\{1\geq w_{1}\geq w_{2}\geq w_{3}\geq0\}$, and $Q(S,T)=-S w_{1}(w_{2}-w_{3})
-T (w_{1}-w_{2})(1-w_{3})$ where $S=-(k_{1}+k_{2})^2$, $T=-(k_{1}+k_{3})^2$
and $U=-(k_{1}+k_{4})^2$. For compactification on a square torus we have
\begin{equation}
G^{IJ}m_{I}m_{J}= \left(m_{1}\over R_{10}\right)^2 +  \left(m_{2}\over R_{11}\right)^2\ .
\end{equation}
In the zero Kaluza-Klein sector $m_{1}=m_{2}=0$ this integral has the $(-S)^{1/2}$
non-analytic behavior of the scalar field theory box diagram in nine-dimensions
\begin{equation}
I_{Box}^0(S,T) \sim   \int_{\mathcal{T}_{ST}} Q(S,T)^{\half}\ .
\end{equation}

In \cite{Russo:1997mk,gkv:twoloop} it was shown that the
derivative expansion of the nine-dimensional expression for the box diagram is given by
\begin{eqnarray}
I_{Box}(s,t)&=& 2\zeta(3) \, e^{-2\phi^{A}}+ {2\pi^2\over3 r_{A}^2}+
 {2\pi^2\over3}
- 8\pi^2\, r_{A}^{-1}\, (l_{s}^2\hat D^2)^{1\over2}\\
\nn&+&8\pi^{3\over2}\sum_{n\geq2} {\Gamma(n-\half)\over n!}\, \zeta(2n-1)\,
r_{A}^{2n-2} (-l_{s}^2\hat D^2)^n\\
\nn&+&8\pi^{2}\sum_{n\geq2} {\Gamma(n-1)\over n!}\, \zeta(2n-2)\, e^{2(n-1)\phi^{A}}
 (-l_{s}^2\hat D^2)^n\\
\nn&+& \textrm{non-perturbative}
\end{eqnarray}
where
$\hat D^{2n}=\int_{\mathcal{T}_{ST}} \,Q(s,t)^n$.
The infinite series of powers of $e^{\phi_A}$ in the last line is just the series of terms in (\ref{genuses}).  The
powers series in powers of $r_A$ is a series that sums to give the thresholds due to massive Kaluza--Klein intermediate
states,
\begin{eqnarray}
\nn r_{A}^{-1} (l_{s}^2 \hat D^2)^{1\over2}\, \hR^4&-&\pi^{-{1\over2}}\,
\sum_{k=2}^\infty
  {\Gamma(k-\half)\over k!}\,  \zeta(2k-1)\,r_A^{2(k-1)}\,(-l_{s}^2\, \hat D^2)^{k}\hR^4\\
  & =&r^{-1}_{A}\, \hR^4\, \int_{\mathcal{T}_{ST}} \,\sum_{n\in\ZZ}
  \left( \frac{n^2}{r_A^2}+l_{s}^2 Q(s,t)\right)^{\half}\, .
\label{powersappne}
\end{eqnarray}

In  \cite{gkv:twoloop} it was shown that the series on the right-hand side of (\ref{powersappne})
sums to give the $s \ln(-\al s)$ threshold term in the ten-dimensional $r_{A}=R_{10}\sqrt{R_{11}}\to\infty$ limit.
This can also be seen directly from the expression for
$I_{Box}(S,T)$, as follows.
In the the limit of  decompactification to the ten-dimensional IIA  theory
 the scalar box integral is dominated by the sector with $m_{2}=0$ and becomes
 \begin{eqnarray}
 \nn I_{Box}(S,T)&\to&{2\pi^9\over R^{1\over2}_{11}}\,  \int_{0}^\infty {dt\over t^{3\over2}}
\, \int_{\mathcal{T}_{ST}}\, \prod_{r=1}^3 dw_{r}\,
{1\over r_{A}}\sum_{m_1\in \ZZ}\, e^{- (m_{1}/r_{A})^2 \, R_{11}^2\, t- Q(S,T)\, t}\\
&\to&{2\pi^{11\over2}\over R_{11}}\,  \int_{0}^\infty {dt\over t^2}
\, \int_{\mathcal{T}_{ST}}\, \prod_{r=1}^3 dw_{r}\,
 e^{- Q(S,T)\, t}\ .
 \end{eqnarray}
The last expression has  the $s\,\ln(-\al s)$  non-analytic behavior of the scalar
box diagram in ten-dimensions,
\begin{equation}
I_{Box}^0(S,T) \sim   \int_{\mathcal{T}_{ST}} Q(S,T) \ln(Q(S,T))\ .
\end{equation}

(ii)   The second example comes from considering the scalar triangle diagram (figure~\ref{fig:s4logs})
compactified to nine dimensions.  This is given by
\begin{eqnarray}\label{e:triangle}
 I_{Triang}(S) ={\pi^{9\over2}\over l_{P}^2 \mathcal{V}}
 \sum_{m_{I}=(m_{1},m_{2})}\, \int_{0}^\infty{ dt \over t^{\fiveh}} \,
 \int_{0}^1dw_{2}\int_{0}^{w_{2}} dw_{1}\,
 e^{-G^{IJ}m_{I}m_{J}\,t -Q(S)\, t }\ ,
\end{eqnarray}
where $Q(S)=-S\, (1-w_{2})(w_{2}-w_{1})$.
This can be expanded in an infinite power series in $S$, which is given by
\begin{eqnarray}\label{e:triangleexp}
I_{Triang}(S)&=&{2\pi^3\over5}\, \hLambda^5+ {3\pi\over 2} \, \zeta(5)\, e^{-2\phi^{A}}
+ 2\zeta(4)\, {e^{2\phi^{A}}\over r_{A}^4}-
 {4\pi^5\over3} \,(l_{s}^2 \hat D^2)^{3\over2}\\
\nn&+&
  2\pi^{9\over2} \sum_{n\geq1} \,
 {\Gamma(n-\threeh)\over n!}\, \zeta(2n-3)\, r_{A}^{2n-4}\,
  (-l_{s}^2\, \hat D^2)^n\\
  \nn&+&4\pi^5 \sum_{n\geq 1} {\Gamma(n-2)\over n!}\, \zeta(2n-4)\, e^{2(n-1)\phi{A}}
  \, (-l_{s}^2 \hat D^2)^n\\
\nn&+& \textrm{non-perturbative}
\end{eqnarray}
where
$\hat D^{2n}=\int_{1\geq w_{2}\geq w_{1}\geq0} \, Q(s)^n$.
The second line is the infinite series of positive powers of $r_A$ in (\ref{powersapp}) that sum to give the
 nine-dimensional Kaluza--Klein thresholds.  The series of higher genus terms in the third line of (\ref{e:triangleexp})
 corresponds to terms that are increasing powers of $R_{11}$.  These are the terms that resum in the limit
 $R_{11} \to \infty$ to reproduce the $(-S)^{3/2}$ threshold term of the  eleven-dimensional theory.

The fact that the sum of Kaluza--Klein thresholds gives the ten-dimensional threshold behaviour
can be seen by directly analyzing $I_{Triang}(S)$ (\ref{e:triangle}) in the $r_A\to \infty$ limit.
The decompactification limit to ten-dimensional perturbative superstring gives
\begin{eqnarray}
\nn I_{Triang}(S)&\to& {\pi^{9\over2}\over l_{P}^2 R^{1\over2}_{11}}
\, \int_{0}^\infty{ dt \over t^{\fiveh}} \,
 \int_{0}^1dw_{2}\int_{0}^{w_{2}} dw_{1}\,{1\over r_{A}} \sum_{m_{1}\in\ZZ}\,
 e^{-(m_{1}/r_{A})^2 \, R_{11}^2\,t -Q(S)\, t }\\
\nn &\to&  {\pi^5\over l_{P}^2 R_{11}}
\, \int_{0}^\infty{ dt \over t^3} \,
 \int_{0}^1dw_{2}\int_{0}^{w_{2}} dw_{1}\,
 e^{-Q(S)\, t }\\
 &\sim&{\pi^5\over l_{P}^2 R_{11}}   \, \int_{0}^1dw_{2}\int_{0}^{w_{2}} dw_{1}\, Q(S)^2 \, \ln(Q(S))  \ ,
\end{eqnarray}
which has the $S^2\, \ln(-S)$ behaviour needed for  obtaining the $S^4\, \ln(-S)$ non-analytic contributions
at one and two loops in string theory in ten dimensions.


\begin{thebibliography}{99}

\bibitem{Witten:1995ex}
  E.~Witten,
  {\sl String theory dynamics in various dimensions,}
  Nucl.\ Phys.\ B {\bf 443} (1995) 85
  [arXiv:hep-th/9503124].

 \bibitem{Schwarz:1995jq}
  J.~H.~Schwarz,
 {\sl The power of M theory,}
  Phys.\ Lett.\ B {\bf 367} (1996) 97
  [arXiv:hep-th/9510086].

  \bibitem{Aspinwall:1995fw}
  P.~S.~Aspinwall,
  {\sl Some relationships between dualities in string theory,}
  Nucl.\ Phys.\ Proc.\ Suppl.\  {\bf 46} (1996) 30
  [arXiv:hep-th/9508154].

 \bibitem{ggv:oneloop}
  M.~B.~Green, M.~Gutperle and P.~Vanhove,
  {\sl One loop in eleven dimensions,}
  Phys.\ Lett.\ B {\bf 409} (1997) 177
  [arXiv:hep-th/9706175].

\bibitem{gv:D6R4}
  M.~B.~Green and P.~Vanhove,
  {\sl Duality and higher derivative terms in M theory,}
  JHEP {\bf 0601}, 093 (2006)
  [arXiv:hep-th/0510027].
 \bibitem{Russo:1997mk}
  J.~G.~Russo and A.~A.~Tseytlin,
  {\sl One-loop four-graviton amplitude in eleven-dimensional supergravity,}
  Nucl.\ Phys.\ B {\bf 508} (1997) 245
  [arXiv:hep-th/9707134].

  \bibitem{gkv:twoloop}
  M.~B.~Green, H.~h.~Kwon and P.~Vanhove,
 {\sl Two loops in eleven dimensions,}
  Phys.\ Rev.\ D {\bf 61} (2000) 104010
  [arXiv:hep-th/9910055].


\bibitem{Bern:1998ug}
  Z.~Bern, L.~J.~Dixon, D.~C.~Dunbar, M.~Perelstein and J.~S.~Rozowsky,
  {\sl On the relationship between Yang-Mills theory and gravity and its
  implication for ultraviolet divergences,}
  Nucl.\ Phys.\ B {\bf 530} (1998) 401
  [arXiv:hep-th/9802162].



\bibitem{gg:dinstantons}
  M.~B.~Green and M.~Gutperle,
  {\sl Effects of D-instantons,}
  Nucl.\ Phys.\ B {\bf 498} (1997) 195
  [arXiv:hep-th/9701093].

\bibitem{Green:1998by}
  M.~B.~Green and S.~Sethi,
  {\sl Supersymmetry constraints on type IIB supergravity,}
  Phys.\ Rev.\ D {\bf 59} (1999) 046006
  [arXiv:hep-th/9808061].




\bibitem{gv:stringloop}  M.B. Green and  P. Vanhove,
{\sl The low energy expansion of the one-loop type II superstring amplitude},
Phys. Rev. {\bf D 61}, 104011 (2000) [arXiv:hep-th/9910056].



\bibitem{grv:stringloop} M.B.~Green,  J.G.~Russo, P.~Vanhove,
{\sl The one-loop type II four-graviton effective action}, to appear.


\bibitem{Berkovits:2006vc}
  N.~Berkovits,
  {\sl New higher-derivative R**4 theorems,}
  arXiv:hep-th/0609006.

\bibitem{DhokerPhongTwoloop}
E.~D'Hoker and D.~H.~Phong,
  {\sl Lectures on two-loop superstrings,}
  arXiv:hep-th/0211111;\hfill\\
  E.~D'Hoker and D.~H.~Phong,
{\sl Two-loop superstrings. VI: Non-renormalization theorems and the
4-point function,}
  Nucl.\ Phys.\ B {\bf 715}, 3 (2005)
  [arXiv:hep-th/0501197].


  \bibitem{DhokerPhongGutperleTwoloop}
E.~D'Hoker, M.~Gutperle and D.~H.~Phong,
 {\sl Two-loop superstrings and S-duality,}
  Nucl.\ Phys.\ B {\bf 722} (2005) 81
  [arXiv:hep-th/0503180].

\bibitem{berkovitstwoloop}  N.~Berkovits and C.~R.~Mafra,
 {\sl Equivalence of two-loop superstring amplitudes in the pure spinor and RNS
  formalisms,}
  Phys.\ Rev.\ Lett.\  {\bf 96}, 011602 (2006)
  [arXiv:hep-th/0509234];\break
  N.~Berkovits,
 {\sl Super-Poincare covariant two-loop superstring amplitudes,}
  JHEP {\bf 0601}, 005 (2006)
  [arXiv:hep-th/0503197].



\bibitem{grv:Poisson} M.B.~Green, J.G.~Russo, P.~Vanhove, {\sl Type IIB couplings and Poisson Equation}, in preparation.

\bibitem{Green:1982sw}
  M.~B.~Green, J.~H.~Schwarz and L.~Brink,
  {\sl N=4 Yang-Mills And N=8 Supergravity As Limits Of String Theories},
  Nucl.\ Phys.\ B {\bf 198} (1982) 474.






\bibitem{Howe:2002ui}
  P.~S.~Howe and K.~S.~Stelle,
 {\sl Supersymmetry counterterms revisited,}
  Phys.\ Lett.\ B {\bf 554} (2003) 190
  [arXiv:hep-th/0211279].

\bibitem{Smirnov:1999gc}
  V.~A.~Smirnov,
  {\sl Analytical result for dimensionally regularized massless on-shell  double box,}
  Phys.\ Lett.\ B {\bf 460} (1999) 397
  [arXiv:hep-ph/9905323].

\bibitem{Tausk:1999vh}
  J.~B.~Tausk,
  {\sl Non-planar massless two-loop Feynman diagrams with four on-shell legs,}
  Phys.\ Lett.\ B {\bf 469} (1999) 225
  [arXiv:hep-ph/9909506].

  \bibitem{Anastasiou:2000kp}
  C.~Anastasiou, J.~B.~Tausk and M.~E.~Tejeda-Yeomans,
  {\sl The on-shell massless planar double box diagram with an irreducible
  numerator,}
  Nucl.\ Phys.\ Proc.\ Suppl.\  {\bf 89} (2000) 262
  [arXiv:hep-ph/0005328].



\bibitem{Kawai:1985xq}
  H.~Kawai, D.~C.~Lewellen and S.~H.~H.~Tye,
{\sl A Relation Between Tree Amplitudes Of Closed And Open Strings,}
  Nucl.\ Phys.\ B {\bf 269} (1986) 1.



  \bibitem{Bern:2005bb}
  Z.~Bern, N.~E.~J.~Bjerrum-Bohr and D.~C.~Dunbar,
  {\sl Inherited twistor-space structure of gravity loop amplitudes,}
  JHEP {\bf 0505}, 056 (2005)
  [arXiv:hep-th/0501137].

 \bibitem{Bjerrum-Bohr:2005xx}
  N.~E.~J.~Bjerrum-Bohr, D.~C.~Dunbar and H.~Ita,
  {\sl Six-point one-loop N = 8 supergravity NMHV amplitudes and their IR
  behaviour,}
  Phys.\ Lett.\ B {\bf 621}, 183 (2005)
  [arXiv:hep-th/0503102].

\bibitem{Bjerrum-Bohr:2006yw}
  N.~E.~J.~Bjerrum-Bohr, D.~C.~Dunbar, H.~Ita, W.~B.~Perkins and K.~Risager,
 {\sl The no-triangle hypothesis for N = 8 supergravity,}
  arXiv:hep-th/0610043.

\bibitem{Bern:2006kd}
  Z.~Bern, L.~J.~Dixon and R.~Roiban,
 {\sl Is N = 8 supergravity ultraviolet finite?,}
  arXiv:hep-th/0611086.

\bibitem{GreenSchwarzloop}   M.~B.~Green and J.~H.~Schwarz,
 {\sl Supersymmetrical Dual String Theory. 3. Loops And Renormalization,}
   Nucl.\ Phys.\ B {\bf 198}, 441 (1982).

\bibitem{GreenSchwarzWitten} M.B.~Green, J.H.~Schwarz and E.~Witten
{\sl Superstring Theory}, Cambridge, Uk: Univ. Pr. ( 1987)
 ( Cambridge Monographs On Mathematical Physics)

 \bibitem{DhokerPhongRevue}   E.~D'Hoker and D.~H.~Phong,
 { \sl The Geometry Of String Perturbation Theory,}
  Rev.\ Mod.\ Phys.\  {\bf 60} (1988) 917.

\end{thebibliography}
\end{document}